\documentclass[twoside,12pt]{article}%
\usepackage{latexsym}
\usepackage{amsmath}
\usepackage{amsfonts}
\usepackage{amssymb}
\usepackage{graphicx}%
\newtheorem{theorem}{Theorem}

\newtheorem{lemma}{Lemma}

\newtheorem{proposition}{Proposition}

\begin{document}
\input{amssym.def}

\thispagestyle{empty}

\begin{center}
\noindent{\large {\textbf{MODEL-ROBUST\ DESIGNS\ FOR\ QUANTILE\ REGRESSION}}%
}\bigskip

\noindent{\textbf{{\large Linglong Kong and Douglas P. Wiens}}}%
\footnote{\noindent{{\footnotesize Department of Mathematical and Statistical
Sciences; University of Alberta, Edmonton, Alberta; Canada T6G 2G1. e-mail:
\texttt{lkong@ualberta.ca, doug.wiens@ualberta.ca}} }}{\textbf{{\large \ }}%
}$\medskip$

\noindent{\today}
\end{center}

\medskip

\noindent{\textbf{Abstract \ }}{We give methods for the construction of
designs for regression models, when the purpose of the investigation is the
estimation of the conditional quantile function and the estimation method is
quantile regression. The designs are robust against misspecified response
functions, and against unanticipated heteroscedasticity. The methods are
illustrated by example, and in a case study} in which they are applied to
growth charts. \medskip

 \medskip Uniformity


\section{Introduction\label{section: intro}}

The need for robust methods of analysis in statistical investigations was
convincingly made by Huber (1981), in whose work one finds a concentration on
robustness against departures from the investigator's assumed parametric model
of the \textit{distribution} generating the data. Box and Draper (1959) had
earlier made the case that, when there is any doubt about the form of the
\textit{response model} in a regression analysis in which the choice of design
is under the control of the experimenter, then such choices should be made
robustly, i.e.\ with an eye to the performance of the resulting designs under
a range of plausible alternate models. A focus of the work of Box and Draper
was on designs robust against polynomial responses of degrees higher than that
anticipated by the experimenter. This was extended in one direction by Huber
(1975), who derived \textit{minimax} designs for straight line fits; these
minimize the maximum mean squared error of the fitted values, with the maximum
taken over a full $L_{2}$-neighbourhood of the experimenter's assumed
response. This work, for which it was assumed that the regression estimates
would be obtained by least squares, has in turn been extended in numerous
directions -- Li (1984) to finite design spaces, Wiens (1992) to multiple
regression, Woods, Lewis, Eccleston and Russell (2006) to GLMs, Li and Wiens
(2011) to dose-response studies, to list but a few.

A method of estimation with a degree of \textit{distributional robustness} is
M-estimation (Huber 1964). Such methods convey robustness against outliers in
the response variable of a regression, but have influence functions which are
unbounded in the factor space. For random regressors this unboundedness may be
addressed by the use of Bounded Influence (BI) methods (Maronna and Yohai
1981, Simpson, Ruppert and Carroll 1992); otherwise it can be controlled by
the design. Designs to be used in league with M- or BI-estimates have been
studied by Wiens (2000) and Wiens and Wu (2010). In the latter article it was
found that there is very little difference between designs optimal (in some
sense, robust or not) for least squares and those for M-estimation; this is
however not the case for BI-estimation.

An increasingly popular method of estimation and inference was furnished by
Koenker and Bassett (1978), who elegantly restated the case for robustness,
went on to extend the notion of univariate quantiles to regression quantiles,
and derived \textit{quantile regression} methods of estimating the conditional
quantile function. Koenker and Bassett point out that the influence function
of a quantile regression estimator is, like that of an M-estimator, unbounded
in the factor space. This can again be addressed by the design. Dette and
Trampisch (2012) have recently studied this problem, assuming that the
experimenter's assumed model is correct; to date there is no published work on
designs for quantile regression methods, which extends the natural robustness
of these methods against outliers to robustness against misspecified response
models. We do so in this article, and also consider robustness against
unanticipated heteroscedasticity.

The need for optimal designs for quantile regression methods was convincingly
articulated by Dette and Trampisch (2012). That for robustness of\ design can
arise in numerous ways. Beyond the obvious -- that in many studies the fitted
model is adopted largely as an article of faith -- there are numerous
scenarios in which the final goal is to fit models which might not fall within
a standard design paradigm, but for which a preliminary study with reasonable
efficiency against a range of models might furnish a point from which to
expand the investigations. Some recent examples of this employ quantile
regression in model selection (Behl, Claeskes and Dette 2014), ecological
studies (Mart\'{\i}nez-Silva, Roca-Pardi\~{n}as, Lustres-P\'{e}rez,
Lorenzo-Arribas and Cadaro-Su\'{a}rez 2013), financial modelling (Rubia and
Sanchis-Marco 2013), and the fitting of time-varying coefficients (Ma and Wei
2012). In such cases the initial fitted model might be non-linear; we address
this in \S 2.2.

In \S \ref{section: approx models} we outline our notion of misspecified
response models, and set the stage for the optimality problems to be addressed
in subsequent sections. The misspecification engenders a bias in the estimate,
motivating our use of mean squared error (\textsc{mse}) of the estimate of the
conditional quantile function as a measure of the loss. In
\S \ref{section: opt var fixed} we illustrate some designs which minimize the
maximum \textsc{mse}, with this maximum taken over certain very broad classes
of response misspecifications. Then in \S \ref{section: minimum bias designs}
we specialize to designs which address only the bias component of the
\textsc{mse} -- this is somewhat of a return to the findings of Box and
Draper, who state (Box and Draper 1959, p. 622) that `... the optimal design
in typical situations in which both variance and bias occur is very nearly the
same as would be obtained if \textit{variance were ignored completely} and the
experiment designed so as to \textit{minimize the bias alone}.'\ 

The bias-minimizing designs turn out to have design weights proportional to
the square roots of the variance functions, when these functions are known. If
instead they are also allowed to range over a certain broad class of variance
functions, then \textit{uniform} designs minimize the maximum bias over both
types of departures from the experimenter's assumptions. In
\S \ref{section: minimum mse designs} this optimality of uniform designs is
extended to minimization of the maximum \textsc{mse} over both types of
departures; we find that the minimax designs are uniform on their support.
Finally, in \S \ref{section: case study}, we illustrate the theory we have
developed in an application to growth charts, in which the regressors are
cubic B-splines and the appropriate choice of knots, and their locations, is
in doubt.

We have posted software (see http://www.stat.ualberta.ca/\symbol{126}%
wiens/home page/pubs/qrd.zip) which runs on \textsc{matlab}, and instructions
for its use, to compute the optimal designs in all of these scenarios. All
derivations and longer mathematical arguments are in the Appendix, or in the
online addendum Kong and Wiens (2014).

\section{Approximate quantile regression models\label{section: approx models}}

To set the stage for the examples of subsequent sections, suppose that an
experimenter intends to make observations on random variables $Y$ with
structure
\begin{equation}
Y=\boldsymbol{f}^{\prime}\left(  \boldsymbol{x}\right)  \boldsymbol{\theta
}+\sigma\left(  \boldsymbol{x}\right)  \varepsilon, \label{approx model}%
\end{equation}
for a $p$-vector $\boldsymbol{f}$ of functionally independent regressors, each
element of which is a function of a $q$-vector $\boldsymbol{x}$ of independent
variables chosen (the `design') from a space $\chi$. We assume that the errors
$\varepsilon$ are i.i.d., and that the variance function $\sigma^{2}\left(
\boldsymbol{x}\right)  $ is strictly positive on the support of the design.
For a fixed $\tau\in\left(  0,1\right)  $, $\boldsymbol{f}^{\prime}\left(
\boldsymbol{x}\right)  \boldsymbol{\theta}$ is to be the conditional $\tau
$-quantile of $Y$, given $\boldsymbol{x}$:
\begin{equation}
\tau=G_{\varepsilon}\left(  0\right)  =G_{Y|\boldsymbol{x}}\left(
\boldsymbol{f}^{\prime}\left(  \boldsymbol{x}\right)  \boldsymbol{\theta
}\right)  . \label{tau}%
\end{equation}
(We write $G_{U}\left(  \cdot\right)  $ for the distribution function of a
random variable $U$.)

Now suppose that (\ref{approx model}) is only an approximation, and that in
fact
\begin{equation}
Y=\boldsymbol{f}^{\prime}\left(  \boldsymbol{x}\right)  \boldsymbol{\theta
}+\delta_{n}\left(  \boldsymbol{x}\right)  +\sigma\left(  \boldsymbol{x}%
\right)  \varepsilon, \label{true model}%
\end{equation}
for some `small' model error $\delta_{n}$. The dependence of $\delta$ on $n$
is necessary for a sensible asymptotic treatment -- in order that bias and
variance remain of the same order we will assume that $\delta_{n}=O\left(
n^{-1/2}\right)  $. For fixed sample sizes this is moot.

The experimenter, acting as though $\delta_{n}\equiv0$ and $\sigma\left(
\cdot\right)  $ is constant, computes the quantile regression estimate
\begin{equation}
\boldsymbol{\hat{\theta}}=\arg\min_{\boldsymbol{t}}\sum_{i=1}^{n}\rho_{\tau
}\left(  Y_{i}-\boldsymbol{f}^{\prime}\left(  \boldsymbol{x}_{i}\right)
\boldsymbol{t}\right)  , \label{thetahat}%
\end{equation}
where $\rho_{\tau}\left(  \cdot\right)  $ is the `check' function $\rho_{\tau
}\left(  r\right)  =r\left(  \tau-I\left(  r<0\right)  \right)  $, with
derivative $\psi_{\tau}\left(  r\right)  =\tau-I\left(  r<0\right)  $.

We will consider two types of design spaces $\chi$. The first is discrete,
with $N$ possible design points $\left\{  \boldsymbol{x}_{i}\right\}
_{i=1}^{N}$; here $N$ is arbitrary. We also consider a continuous, compact
design space, with Lebesgue measure \textsc{vol}$\left(  \chi\right)
\overset{def}{=}\int_{\chi}d\boldsymbol{x}$, in which case the design is
generated by a design measure $\xi\left(  d\boldsymbol{x}\right)  $.
Initially, we shall unify the presentation by writing sums of the form
$\sum_{\boldsymbol{x}\in design}\alpha\left(  \boldsymbol{x}\right)  $, in
which a fraction $\xi_{n,i}=$ $n_{i}/n$ of the $n$ observations are to be made
at the design point $\boldsymbol{x}=\boldsymbol{x}_{i}$, as Lebesgue-Stieltjes
integrals, viz.\ as $n\sum_{i=1}^{N}\xi_{n,i}\alpha\left(  \boldsymbol{x}%
_{i}\right)  =n\int_{\chi}\alpha\left(  \boldsymbol{x}\right)  \xi_{n}\left(
d\boldsymbol{x}\right)  $. We assume that the design measure $\xi_{n}$ has a
weak limit $\xi_{\infty}$ for which%
\[
\lim_{n\rightarrow\infty}\sum_{i=1}^{n}\xi_{n,i}\rho_{\tau}\left(
Y_{i}-\boldsymbol{f}^{\prime}\left(  \boldsymbol{x}_{i}\right)  \boldsymbol{t}%
\right)  =\int_{\chi}E_{Y|\boldsymbol{x}}\left[  \rho_{\tau}\left(
Y-\boldsymbol{f}^{\prime}\left(  \boldsymbol{x}\right)  \boldsymbol{t}\right)
\right]  \xi_{\infty}\left(  d\boldsymbol{x}\right)  .
\]

Under (\ref{true model}) the meaning of $\boldsymbol{\theta}$ becomes
ambiguous. Thus we define this `true' regression parameter as that making the
experimenter's approximation (\ref{approx model}) most accurate, under the
experimenter's assumption of homoscedasticity. For a discrete design space
this is%
\begin{equation}
\boldsymbol{\theta}=\arg\min_{\boldsymbol{t}}\frac{1}{N}\sum_{i=1}%
^{N}E_{Y|\boldsymbol{x}}\left[  \rho_{\tau}\left(  Y-\boldsymbol{f}^{\prime
}\left(  \boldsymbol{x}_{i}\right)  \boldsymbol{t}\right)  \right]
.\label{true theta}%
\end{equation}
Carrying out the minimization in (\ref{true theta}) and evaluating at
$\boldsymbol{t}=\boldsymbol{\theta}$:
\begin{align}
\boldsymbol{0} &  =\frac{1}{N}\sum_{i=1}^{N}E_{Y|\boldsymbol{x}}\left[
\psi_{\tau}\left(  Y-\boldsymbol{f}^{\prime}\left(  \boldsymbol{x}_{i}\right)
\boldsymbol{t}\right)  \right]  \boldsymbol{f}\left(  \boldsymbol{x}%
_{i}\right)  \label{root1}\\
&  =\frac{1}{N}\sum_{i=1}^{N}\left[  G_{\varepsilon}\left(  0\right)
-G_{\varepsilon}\left(  -\delta_{n}\left(  \boldsymbol{x}_{i}\right)  \right)
\right]  \boldsymbol{f}\left(  \boldsymbol{x}_{i}\right)  \nonumber\\
&  =\left(  g_{\varepsilon}\left(  0\right)  +O\left(  n^{-1/2}\right)
\right)  \frac{1}{N}\sum_{i=1}^{N}\delta_{n}\left(  \boldsymbol{x}_{i}\right)
\boldsymbol{f}\left(  \boldsymbol{x}_{i}\right)  ,\nonumber
\end{align}
where $g_{\varepsilon}$ is the density of $G_{\varepsilon}$. We now define
$\delta_{0}(\boldsymbol{x})=\lim_{n\rightarrow\infty}\sqrt{n}\delta
_{n}(\boldsymbol{x})$, so that%
\begin{equation}
\frac{1}{N}\sum_{i=1}^{N}\delta_{0}\left(  \boldsymbol{x}_{i}\right)
\boldsymbol{f}\left(  \boldsymbol{x}_{i}\right)  =\boldsymbol{0}%
.\label{orthogonality}%
\end{equation}
In a continuous design space the average is replaced by an integral -- see
(\ref{Delta0continuous}).

The true conditional $\tau$-quantile $Y_{\tau}=\boldsymbol{f}^{\prime}\left(
\boldsymbol{x}\right)  \boldsymbol{\theta}+\delta_{n}\left(  \boldsymbol{x}%
\right)  $ is predicted by $\hat{Y}_{\tau}=\boldsymbol{f}^{\prime}\left(
\boldsymbol{x}\right)  \boldsymbol{\hat{\theta}}$, and our approach is to
obtain the asymptotic mean squared error matrix \textsc{mse}%
$_{\boldsymbol{\hat{\theta}}}$ of the parameter estimates, thus obtaining the
average -- over $\chi$ -- \textsc{mse} of these predicted values, and to
maximize this average \textsc{mse} over the appropriate choice
\begin{subequations}
\label{Delta0}%
\begin{align}
&  \chi\text{ discrete}\text{:\ }\Delta_{0}=\left\{  \delta_{0}\left(
\boldsymbol{\cdot}\right)  \left\vert {}\right.  \text{ (i) }N^{-1}\sum
_{i=1}^{N}\text{ }\delta_{0}\left(  \boldsymbol{x}_{i}\right)  \boldsymbol{f}%
\left(  \boldsymbol{x}_{i}\right)  =\boldsymbol{0}\ \text{and (ii) }N^{-1}%
\sum_{i=1}^{N}\delta_{0}^{2}\left(  \boldsymbol{x}_{i}\right)  \leq\eta
^{2}\right\}  ,\label{Delta0discrete}\\
&  \chi\text{ continuous}\text{:\ }\Delta_{0}=\left\{  \delta_{0}\left(
\boldsymbol{\cdot}\right)  \left\vert {}\right.  \text{ (i) }\int_{\chi}%
\delta_{0}\left(  \boldsymbol{x}\right)  \boldsymbol{f}\left(  \boldsymbol{x}%
\right)  d\boldsymbol{x}=\boldsymbol{0}\ \text{and (ii) }\int_{\chi}\delta
_{0}^{2}\left(  \boldsymbol{x}\right)  d\boldsymbol{x}\leq\eta^{2}\right\}
.\label{Delta0continuous}%
\end{align}

This is carried out in \S \ref{section: max amse delta}. We also consider
classes of variance functions. These may be independent of the design or --
see (\ref{var funcs}) -- vary with the designs weights, in which case we also
maximize the \textsc{mse} over this class. In any event we then go on to find
the \textsc{mse}\textit{-}minimizing designs $\xi_{\ast}$, using a variety of
analytic and numerical techniques.

In most cases the optimal designs $\xi_{\ast}$ must be approximated in order
to implement them in finite samples; for example when $q=1$ we will do this by
placing the design points at the quantiles
\end{subequations}
\begin{equation}
x_{i}=\xi_{\ast}^{-1}\left(  \frac{i-.5}{n}\right)  , \label{implement}%
\end{equation}
or at the closest available points in discrete design spaces. For $q>1$ the
situation is more interesting and some suggestions are in Fang and Wang (1994)
and Xu and Yuen (2011); an intriguing possibility as yet (to our knowledge)
unexplored is the use of vector quantization to approximate the designs.

\subsection{Asymptotics\label{section: asymptotics}}

In (\ref{Delta0}) the imposition of (\ref{orthogonality}), and its analogue in
continuous spaces, ensures the identifiability of the parameter in
(\ref{true model}). The bounds of $\eta^{2}$ force the errors due to
variation, and those due to the bias engendered by the model misspecification,
to remain of the same order asymptotically -- a situation akin to the
imposition of contiguity in the asymptotic theory of hypothesis testing.
Define
\begin{subequations}
\label{denn}%
\begin{align}
\boldsymbol{\mu}_{0}  &  =\int_{\chi}\text{ }\delta_{0}(\boldsymbol{x}%
)\frac{1}{\sigma(\boldsymbol{x})}\boldsymbol{f}(\boldsymbol{x})\xi_{\infty
}\left(  d\boldsymbol{x}\right)  ,\label{defn1}\\
\boldsymbol{P}_{0}  &  =\int_{\chi}\boldsymbol{f}(\boldsymbol{x}%
)\boldsymbol{f}^{\prime}(\boldsymbol{x})\xi_{\infty}\left(  d\boldsymbol{x}%
\right)  ,\label{defn2}\\
\boldsymbol{P}_{1}  &  =\int_{\chi}\boldsymbol{f}(\boldsymbol{x})\frac
{1}{\sigma(\boldsymbol{x})}\boldsymbol{f}^{\prime}(\boldsymbol{x})\xi_{\infty
}\left(  d\boldsymbol{x}\right)  . \label{defn3}%
\end{align}
Assume that the support of $\xi_{\infty}$ is large enough that $\boldsymbol{P}%
_{0}$ and $\boldsymbol{P}_{1}$ are positive definite. Define the target
parameter $\boldsymbol{\theta}$ to be the asymptotic solution to
(\ref{thetahat}), so that
\end{subequations}
\begin{equation}
\sum_{i=1}^{n}\xi_{n,i}\psi_{\tau}\left(  Y_{i}-\boldsymbol{f}^{\prime}\left(
\boldsymbol{x}_{i}\right)  \boldsymbol{\theta}\right)  \boldsymbol{f}\left(
\boldsymbol{x}_{i}\right)  \overset{pr}{\rightarrow}\boldsymbol{0},
\label{target}%
\end{equation}
in agreement with (\ref{root1}). The proof of the asymptotic normality of the
estimate runs along familiar lines -- see Knight (1998) and Koenker (2005) --
and so we merely state the result. Complete details are in Kong and Wiens (2014).

\begin{theorem}
\label{thm: asymptotic normality}Under conditions (A1) -- (A3) the quantile
regression estimate $\boldsymbol{\hat{\theta}}_{n}$ of the parameter
$\boldsymbol{\theta}$ defined by (\ref{target}) is asymptotically normally
distributed:
\[
\sqrt{n}\left(  \boldsymbol{\hat{\theta}}_{n}-\boldsymbol{\theta}\right)
\overset{L}{\rightarrow}N\left(  \boldsymbol{P}_{1}^{-1}\boldsymbol{\mu}%
_{0},\frac{\tau\left(  1-\tau\right)  }{g_{\varepsilon}^{2}\left(  0\right)
}\boldsymbol{P}_{1}^{-1}\boldsymbol{P}_{0}\boldsymbol{P}_{1}^{-1}\right)  .
\]

\end{theorem}

\subsection{\noindent Nonlinear models}

Dette and Trampisch (2012) obtained (non-robust) designs for quantile
regression and \textit{nonlinear} models; these were \textit{locally optimal},
with, in our notation, (\ref{tau}) replaced by $\tau=G_{Y|\boldsymbol{x}%
}\left(  F\left(  \boldsymbol{x};\boldsymbol{\theta}\right)  \right)  $, where
$F\left(  \boldsymbol{x};\boldsymbol{\theta}\right)  $ is evaluated at fixed
values of those elements of $\boldsymbol{\theta}$ which enter in a nonlinear
manner. Some robustness against misspecifications of these local parameters
was then introduced by considering Bayesian and maximin designs. In this
article the examples pertain only to \textit{linear} models. However, since
(as also in Dette and Trampisch 2012), our approach is asymptotic in nature,
the results presented here are easily modified to accommodate nonlinear
models. The definition (\ref{thetahat}) of the estimate is replaced by
$\boldsymbol{\hat{\theta}}=\arg\min_{\boldsymbol{t}}\sum_{i=1}^{n}\rho_{\tau
}\left(  Y_{i}-F\left(  \boldsymbol{x}_{i};\boldsymbol{t}\right)  \right)  $,
and then, in all occurrences, $\boldsymbol{f}\left(  \boldsymbol{x}\right)  $
is to be replaced by the gradient $\boldsymbol{f}_{\boldsymbol{\theta}}\left(
\boldsymbol{x}\right)  =\partial F\left(  \boldsymbol{x};\boldsymbol{\theta
}\right)  /\partial\boldsymbol{\theta}$. With these changes Theorem
\ref{thm: asymptotic normality} continues to hold, as does the rest of the
theory of the article. The robustness is then attained against
misspecifications in the functional form of $F\left(  \boldsymbol{x}%
;\boldsymbol{\cdot}\right)  $, possibly but not necessarily arising from
misspecified parameters.

\subsection{Maximum MSE over $\Delta_{0}$; discrete design spaces
\label{section: max amse delta}}

From Theorem \ref{thm: asymptotic normality}, the asymptotic \textsc{mse}
matrix\ of $\boldsymbol{\hat{\theta}}$ is%
\[
\text{\textsc{mse}}_{\boldsymbol{\hat{\theta}}}=\boldsymbol{P}_{1}^{-1}\left[
\frac{\tau\left(  1-\tau\right)  }{g_{\varepsilon}^{2}\left(  0\right)
}\boldsymbol{P}_{0}+\boldsymbol{\mu}_{0}\boldsymbol{\mu}_{0}^{\prime}\right]
\boldsymbol{P}_{1}^{-1}.
\]
We now introduce a measure of the asymptotic loss when the conditional
quantile $Y_{\tau}\left(  \boldsymbol{x}\right)  =\boldsymbol{f}^{\prime
}\left(  \boldsymbol{x}\right)  \boldsymbol{\theta}+\delta_{n}\left(
\boldsymbol{x}\right)  $, for $\boldsymbol{x}\in\chi$, is incorrectly
estimated by $\hat{Y}_{n}\left(  \boldsymbol{x}\right)  =\boldsymbol{f}%
^{\prime}\left(  \boldsymbol{x}\right)  \hat{\boldsymbol{\theta}}_{n}$. For a
discrete design space $\chi=\left\{  \boldsymbol{x}_{1},...,\boldsymbol{x}%
_{N}\right\}  $ this measure is the limiting average mean squared error%
\[
\text{\textsc{amse}}=\lim_{n}\frac{1}{N}\sum_{i=1}^{N}E\left[  \left\{
\sqrt{n}\left(  \hat{Y}_{n}\left(  \boldsymbol{x}_{i}\right)  -Y_{\tau}\left(
\boldsymbol{x}_{i}\right)  \right)  \right\}  ^{2}\right]  .
\]
In terms of $\boldsymbol{A}=N^{-1}\sum_{i=1}^{N}\boldsymbol{f}\left(
\boldsymbol{x}_{i}\right)  \boldsymbol{f}^{\prime}\left(  \boldsymbol{x}%
_{i}\right)  $, and using (i) of (\ref{Delta0discrete}), we find that%
\begin{align}
\text{\textsc{amse}}  &  =tr\left(  \boldsymbol{A\cdot}\text{\textsc{mse}%
}_{\boldsymbol{\hat{\theta}}}\right)  +\frac{1}{N}\sum_{i=1}^{N}\delta_{0}%
^{2}\left(  \boldsymbol{x}_{i}\right) \nonumber\\
&  =\frac{\tau\left(  1-\tau\right)  }{g_{\varepsilon}^{2}\left(  0\right)
}tr\left(  \boldsymbol{AP}_{1}^{-1}\boldsymbol{P}_{0}\boldsymbol{P}_{1}%
^{-1}\right)  +\boldsymbol{\mu}_{0}^{\prime}\boldsymbol{P}_{1}^{-1}%
\boldsymbol{AP}_{1}^{-1}\boldsymbol{\mu}_{0}+\frac{1}{N}\sum_{i=1}^{N}%
\delta_{0}^{2}\left(  \boldsymbol{x}_{i}\right)  . \label{amse}%
\end{align}

We now write merely $\xi$ for $\xi_{\infty}$. We impose a bound $N^{-1}%
\sum_{i=1}^{N}\sigma^{2}(\boldsymbol{x}_{i})\leq\sigma_{0}^{2}$ for a given
$\sigma_{0}^{2}$, and we denote by $ch_{\max}$ the maximum eigenvalue of a
matrix. The maximum value of \textsc{amse} over $\Delta_{0}$ is given in the
following theorem.

\begin{theorem}
\label{thm: max amse discrete}For a discrete design space $\chi$ define%
\[
\boldsymbol{T}_{0,0}=\sum_{\xi_{i}>0}\boldsymbol{f}(\boldsymbol{x}%
_{i})\boldsymbol{f}^{\prime}(\boldsymbol{x}_{i})\xi_{i}\text{, }%
\boldsymbol{T}_{0,k}=\sum_{\xi_{i}>0}\boldsymbol{f}(\boldsymbol{x}%
_{i})\boldsymbol{f}^{\prime}(\boldsymbol{x}_{i})\left(  \frac{\xi_{i}}%
{\sigma(\boldsymbol{x}_{i})/\sigma_{0}}\right)  ^{k},\ k=1,2,
\]
and
\begin{equation}
\boldsymbol{T}_{0}=\boldsymbol{T}_{0,1}^{-1}\boldsymbol{T}_{0,0}%
\boldsymbol{T}_{0,1}^{-1},\ \boldsymbol{T}_{2}=\boldsymbol{T}_{0,1}%
^{-1}\boldsymbol{T}_{0,2}\boldsymbol{T}_{0,1}^{-1}. \label{T1T2}%
\end{equation}
Then $\max_{\Delta_{0}}$\textsc{amse} is $\frac{\tau\left(  1-\tau\right)
\sigma_{0}^{2}}{g_{\varepsilon}^{2}\left(  0\right)  }+\eta^{2}$ times
\begin{equation}
\mathcal{L}_{\nu}\left(  \xi|\sigma\right)  =\left(  1-\nu\right)  tr\left(
\boldsymbol{AT}_{0}\right)  +\nu ch_{\max}\left(  \boldsymbol{AT}_{2}\right)
, \label{max loss|sigma}%
\end{equation}
where $\nu=\eta^{2}\left/  \left\{  \frac{\tau\left(  1-\tau\right)
\sigma_{0}^{2}}{g_{\varepsilon}^{2}\left(  0\right)  }+\eta^{2}\right\}
\right.  $ .
\end{theorem}

The first component ($tr\left(  \boldsymbol{AT}_{0}\right)  $) of
$\mathcal{L}_{\nu}\left(  \xi|\sigma\right)  $ arises solely from variation,
the second ($ch_{\max}\left(  \boldsymbol{AT}_{2}\right)  $) from (squared)
bias. Note that (\ref{max loss|sigma} ) depends on $\sigma_{0}$ only through
$\left\{  \sigma(\boldsymbol{x})/\sigma_{0}\right\}  $ and through $\nu$. We
may thus without loss of generality take $\sigma_{0}=1$ and parameterize the
designs solely by $\nu\in\left[  0,1\right]  $, which may be chosen by the
experimenter, representing his relative concern for errors due to bias rather
than to variation.

\section{Examples: Designs minimizing max$_{\Delta_{0}}$
{\protect\normalsize MSE} for fixed variance
functions\label{section: opt var fixed}}

Before extending the theory presented thus far, we illustrate it for some
representative, fixed variance functions in two cases -- approximate straight
line regression in a discrete design space, and approximate quadratic
regression in a continuous design space. The development of the first case is
given in some detail in the Appendix; that for the second is outlined only briefly.

\subsection{Discrete design spaces\label{section: fixed var discrete}}

For least squares regression problems with univariate design variables and
homoscedastic variances, optimally robust designs have been constructed by,
among others, Fang and Wiens (2000), who computed exact designs by simulated
annealing. Here we construct optimal designs for heteroscedastic quantile
regression problems and also take a different approach to the implementation
-- we obtain exact optimal values $\left\{  \xi_{\ast,i}\right\}  $ and then
implement the designs as at (\ref{implement}).%
\begin{figure}
[tb]
\begin{center}
\includegraphics[
trim=0.000000cm 0.708247cm 0.000000cm 0.713796cm,
height=11.1654cm,
width=14.0826cm
]%
{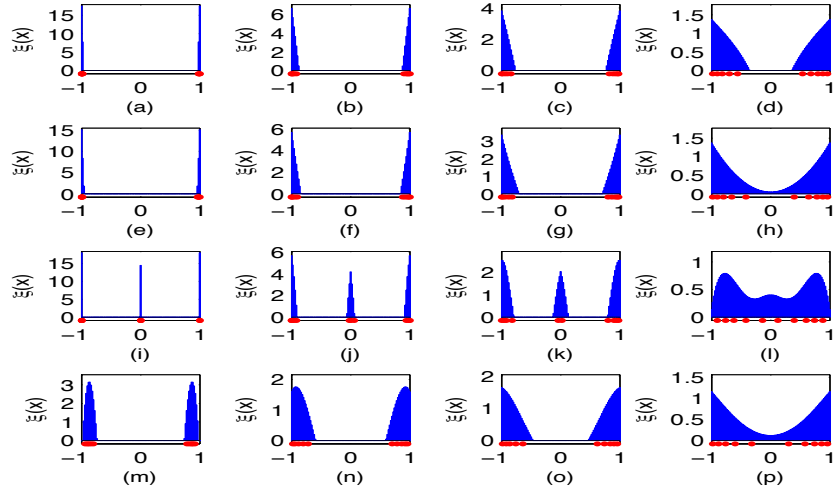}%
\caption{Minimax (over $\Delta_{0}$) design measures for heteroscedastic
straight line regression, $N=101$, normalized so that bar area = 1. Columns 1
-- 4 use $\nu=.05,.35,.65,.95$ respectively; rows 1 -- 4 use $\sigma\left(
x\right)  \propto\left(  1+\left\vert x\right\vert \right)  ^{-1}$, $1$,
$.2+\left\vert x\right\vert $, $1+\left(  x/2\right)  ^{2}$ respectively. The
bullets below the horizontal axes are the locations of $n=10$ design points,
implemented as at (\ref{implement}).}%
\label{fig slr}%
\end{center}
\end{figure}

For a fixed variance function and a discrete design space we seek a design
$\xi_{\ast}$ minimizing (\ref{max loss|sigma}). We illustrate the method in
the case of approximate straight line models -- $\boldsymbol{f}\left(
x_{i}\right)  =\left(  1,x_{i}\right)  ^{\prime}$ -- and suppose that the
design space $\chi$ consists of $N$ points in $\left[  -1,1\right]  $. The
space $\chi$ is symmetric in that if $\chi=\left(  x_{1},...,x_{N}\right)
^{\prime}$ ($-1=x_{1}<\cdot\cdot\cdot<x_{N}=1$) and $\chi_{\pi}$ denotes the
reversal $\left(  x_{N},...,x_{1}\right)  ^{\prime}$ then $\chi_{\pi
}=\boldsymbol{-}\chi$. We consider symmetric designs, i.e.\ designs for which
$\boldsymbol{\xi}=\left(  \xi_{1},...,\xi_{N}\right)  ^{\prime}$, with
$\xi_{i}=\xi\left(  x_{i}\right)  $, satisfies $\xi\left(  x_{i}\right)
=\xi\left(  -x_{i}\right)  $. We also assume a symmetric but arbitrary
variance function $\sigma_{i}=\sigma\left(  \left\vert x_{i}\right\vert
\right)  $.

The designs are obtained by variational arguments followed by a constrained
numerical minimization; the details are in the Appendix. See Figure
\ref{fig slr} for representative plots of the designs, scaled so as to have
unit area. In these plots the bullets below the horizontal axes are the
locations of $n=10$ design points, implemented as at (\ref{implement}). In the
case of homoscedasticity (plots (e) - (h)) the designs for very small $\nu$
are close in nature to their non-robust counterparts, placing point masses at
$\pm1$. As $\nu$ increases these replicates spread out into clusters near
$\pm1$ and, depending upon the variance function, possibly near $0$ as well.
The limiting behaviour as $\nu\rightarrow1$ is studied in
\S \ref{section: minimum bias designs}.

\subsection{Continuous design spaces \label{section: fixed var continuous}}

The continuous case requires special consideration. Rather than \textsc{amse}
at (\ref{amse}) we use instead the integrated mean squared error%
\[
\text{\textsc{imse}}=\lim_{n}\int_{\chi}E\left[  \left\{  \sqrt{n}\left(
\hat{Y}_{n}\left(  \boldsymbol{x}\right)  -Y_{\tau}\left(  \boldsymbol{x}%
\right)  \right)  \right\}  ^{2}\right]  d\boldsymbol{x},
\]
together with $\boldsymbol{A}=\int_{\chi}\boldsymbol{f}(\boldsymbol{x}%
)\boldsymbol{f}^{\prime}(\boldsymbol{x})d\boldsymbol{x}$, and obtain
\[
\text{\textsc{imse}}=\frac{\tau\left(  1-\tau\right)  }{g_{\varepsilon}%
^{2}\left(  0\right)  }tr\left(  \boldsymbol{AP}_{1}^{-1}\boldsymbol{P}%
_{0}\boldsymbol{P}_{1}^{-1}\right)  +\boldsymbol{\mu}_{0}^{\prime
}\boldsymbol{P}_{1}^{-1}\boldsymbol{AP}_{1}^{-1}\boldsymbol{\mu}_{0}%
+\int_{\chi}\delta_{0}^{2}\left(  \boldsymbol{x}\right)  d\boldsymbol{x}.
\]

In order that the maximum \textsc{imse}\ be finite, it is necessary that the
design measure be absolutely continuous. That this should be so is intuitively
clear -- if $\xi_{\infty}$ in (\ref{denn}) places positive mass on sets of
Lebesgue measure zero, such as individual points, then $\delta_{0}$ may be
chosen arbitrarily large on such sets without altering its membership in
$\Delta_{0}$, and one can do this in such a way as to drive \textsc{imse}%
\ beyond all bounds, through (\ref{defn1}). A formal proof may be based on
that of Lemma 1 in Heo, Schmuland and Wiens (2001).

When implementing continuous designs we discretize; for instance when there is
only one covariate we employ (\ref{implement}). As a referee has pointed out,
this might result in an unbounded \textsc{imse} along particularly
pathological sequences $\left\{  \delta_{n}\right\}  $. A possible
alternative, which we do not illustrate here since it is unlikely to find
favour with practitioners, is to randomly choose design points from the
optimal design measure; in the parlance of game theory this would thwart the
intentions of a malevolent Nature, which can then not anticipate the design.

In the same vein Bischoff (2010) states a criticism, in a context of
discretized, absolutely continuous, lack-of-fit designs as proposed by Wiens
(1991) and Biedermann and Dette (2001), of the very rich class of
alternatives, analogous to (\ref{Delta0continuous}), used by those authors.
Bischoff suggests using a smaller class of alternatives; here we are however
in accord with Wiens (1992), who states `\textit{Our attitude is that an
approximation to a design which is robust against more realistic alternatives
is preferable to an exact solution in a neighbourhood which is unrealistically
sparse.}'\ 

We write $m\left(  \boldsymbol{x}\right)  $ for the density of $\xi$ when
dealing with continuous design spaces and take $\int_{\chi}\sigma
^{2}(\boldsymbol{x})d\boldsymbol{x}\leq\sigma_{0}^{2}$ ($=1$, as in the
discrete case).

\begin{theorem}
\label{thm: max amse continuous}For a continuous design space $\chi$ define
$\boldsymbol{T}_{0}$ and $\boldsymbol{T}_{2}$ as at (\ref{T1T2}), with%
\[
\boldsymbol{T}_{0,0}=\int_{\chi}\boldsymbol{f}(\boldsymbol{x})\boldsymbol{f}%
^{\prime}(\boldsymbol{x})m\left(  \boldsymbol{x}\right)  d\boldsymbol{x}%
\text{\ and }\boldsymbol{T}_{0,k}=\int_{\chi}\boldsymbol{f}(\boldsymbol{x}%
)\boldsymbol{f}^{\prime}(\boldsymbol{x})\left(  \frac{m\left(  \boldsymbol{x}%
\right)  }{\sigma(\boldsymbol{x})/\sigma_{0}}\right)  ^{k}d\boldsymbol{x}%
,\text{ }k=1,2.
\]
Then the maximum \textsc{imse}\ is given by (\ref{max loss|sigma}).
\end{theorem}

%

\begin{figure}
[tb]
\begin{center}
\includegraphics[
trim=0.000000cm 0.467406cm 0.000000cm 0.480971cm,
height=11.1632cm,
width=14.0826cm
]%
{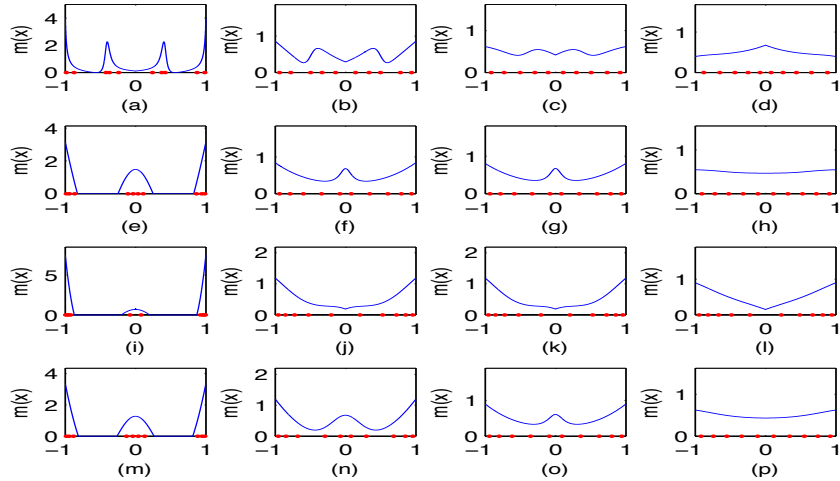}%
\caption{Minimax (over $\Delta_{0}$) design densities for heteroscedastic
quadratic regression on $\left[  -1,1\right]  $. Columns 1 -- 4 use
$\nu=.05,.35,.65,.95$ respectively; rows 1 -- 4 use $\sigma\left(  x\right)
\propto\left(  1+\left\vert x\right\vert \right)  ^{-1}$, $1$, $.2+\left\vert
x\right\vert $, $1+\left(  x/2\right)  ^{2}$ respectively. The bullets on the
horizontal axes are the locations of $n=10$ design points, implemented as at
(\ref{implement}).}%
\label{fig quad}%
\end{center}
\end{figure}

As an example we minimize \textsc{imse} for approximate quadratic regression,
i.e.\ $\boldsymbol{f}(x)=(1,x,x^{2})^{\prime}$, and a fixed variance function
$\sigma^{2}\left(  x\right)  $, over the design space $\chi=[-1,1]$. Similar
problems, assuming homoscedasticity, were studied previously by Shi, Ye and
Zhou (2003) using methods of nonsmooth optimization, and by Daemi and Wiens
(2013) following the methods used here and outlined in the Appendix.

We show in the Appendix that the minimizing density is of the form
\begin{equation}
m(x;\mathbf{a})=\left(  \frac{q_{1}\left(  x\right)  \sigma\left(  x\right)
+q_{2}\left(  x\right)  }{a_{00}+\frac{q_{3}\left(  x\right)  }{\sigma\left(
x\right)  }}\right)  ^{+}, \label{form of m}%
\end{equation}
for polynomials $q_{j}\left(  x\right)  =a_{0j}+a_{2j}x^{2}+a_{4j}x^{4}$,
$j=1,2,3$. The ten constants $a_{ij}$ forming $\mathbf{a}$ are chosen to
minimize the loss $\mathcal{L}_{\nu}\left(  \xi|\sigma\right)  $ at
(\ref{max loss|sigma}) over $\mathbf{a}$, subject to $\int_{-1}^{1}%
m(x;\mathbf{a})dx=1$. Some examples are illustrated in Figure \ref{fig quad}.
Again there is a pronounced increase in the spreading out of the mass as $\nu$
increases, and again under homoscedasticity these masses are initially
concentrated near $\pm1$ and $0$, as in the non-robust case. It is rather
evident from the plots in the rightmost panels of Figure \ref{fig quad} that,
as $\nu\rightarrow1$, the density $m(x;\mathbf{a})$ becomes proportional to
$\sigma\left(  x\right)  $, a phenomenon explained in the following section.

\section{Bias minimizing designs\label{section: minimum bias designs}}

The following result is quite elementary, but since we use it repeatedly we
give it a formal statement and proof.

\begin{proposition}
\label{prop: inequality}(i) Suppose that $\chi$ is discrete, that the function
$p\left(  \boldsymbol{x}\right)  $ is defined on $\chi_{0}\subset\chi$ and
that $\mathbf{M}_{q}\overset{def}{=}\sum_{\boldsymbol{x}_{i}\in\chi_{0}%
}q\left(  \boldsymbol{x}_{i}\right)  \boldsymbol{f}(\boldsymbol{x}%
_{i})\boldsymbol{f}^{\prime}(\boldsymbol{x}_{i})$ exists for $q=p$, $q=p^{2}$
and $q=\mathbf{1}$ ($\mathbf{1}\left(  \boldsymbol{x}_{i}\right)  \equiv1$),
and is invertible for $q=p$ and $q=\mathbf{1}$. Then, under the ordering
`$\succeq$' with respect to positive semidefiniteness,%
\begin{equation}
\mathbf{M}_{p}^{-1}\mathbf{M}_{p^{2}}\mathbf{M}_{p}^{-1}\succeq\mathbf{M}%
_{\mathbf{1}}^{-1}. \label{ineq1}%
\end{equation}
(ii) Suppose that $\chi$ is continuous, that the function $p\left(
\boldsymbol{x}\right)  $ is defined on $\chi_{0}\subset\chi$ and that
$\mathbf{M}_{q}\overset{def}{=}\int_{\chi_{0}}\boldsymbol{f}(\boldsymbol{x}%
)\boldsymbol{f}^{\prime}(\boldsymbol{x})q\left(  \boldsymbol{x}\right)
d\boldsymbol{x}$ exists for $q=p$, $q=p^{2}$ and $q=\mathbf{1}$, and is
invertible for $q=p$ and $q=\mathbf{1}$. Then (\ref{ineq1}) holds.
\end{proposition}

In discrete design spaces we define $\boldsymbol{A}_{\xi}=\sum_{\xi_{i}%
>0}\boldsymbol{f}(\boldsymbol{x}_{i})\boldsymbol{f}^{\prime}(\boldsymbol{x}%
_{i})$. It then follows from Proposition \ref{prop: inequality} that
$\boldsymbol{T}_{2}\succeq\boldsymbol{A}_{\xi}^{-1}$; note as well that
$\boldsymbol{A}_{\xi}^{-1}\succeq\left(  N\boldsymbol{A}\right)  ^{-1}$.
\medskip Together these imply that
\begin{equation}
\mathcal{L}_{\nu=1}\left(  \xi|\sigma\right)  =ch_{\max}\left(  \mathbf{A}%
\boldsymbol{T}_{2}\right)  \geq ch_{\max}\left(  \mathbf{AA}_{\xi}%
^{-1}\right)  \geq1/N. \label{chmax lower bound}%
\end{equation}

Motivated by the remark of Box and Draper (1959) quoted in \S 1 of this
article we note that, if the experimenter seeks robustness only against errors
due to bias (so that $\nu=1$), whether arising from a misspecified response
model or a particular variance function $\sigma^{2}\left(  \cdot\right)  $,
then the maximum bias is minimized by $\xi_{i}=\sigma\left(  \boldsymbol{x}%
_{i}\right)  \left/  \sum_{i=1}^{N}\sigma\left(  \boldsymbol{x}_{i}\right)
\right.  $, since then $\boldsymbol{T}_{2}=\mathbf{A}_{\xi}^{-1}$ and the
lower bound in (\ref{chmax lower bound}) is attained.

Similarly, in a continuous design space the maximum bias is minimized by
$m\left(  \boldsymbol{x}\right)  =\sigma\left(  \boldsymbol{x}\right)  \left/
\int_{\chi}\sigma(\boldsymbol{x})d\boldsymbol{x}\right.  $; for this we use
$\boldsymbol{A}_{m}=\int_{m(\boldsymbol{x})>0}\boldsymbol{f}(\boldsymbol{x}%
)\boldsymbol{f}^{\prime}(\boldsymbol{x})d\boldsymbol{x}$, in place of
$\boldsymbol{A}_{\xi}$ and obtain a lower bound of $1$ in
(\ref{chmax lower bound}).

If the form of the variance function is in doubt, then a minimax approach
dictates taking a further maximum over a class of such functions. We consider
the class $\Sigma_{0}=\left\{  \sigma_{\xi}^{2}(\cdot|r)|r\in\left(
-\infty,\infty\right)  \right\}  $ of variance functions given by
\begin{equation}
\sigma_{\xi}(\boldsymbol{x}|r)=\left\{
\begin{array}
[c]{cc}%
c_{r}\xi^{r/2}\left(  \boldsymbol{x}\right)  I\left(  \xi\left(
\boldsymbol{x}\right)  >0\right)  , & \chi\text{ discrete,}\\
c_{r}m^{r/2}\left(  \boldsymbol{x}\right)  I\left(  m(\boldsymbol{x}%
)>0\right)  , & \chi\text{ continuous;}%
\end{array}
\right.  \label{var funcs}%
\end{equation}
$c_{r}$ is the required constant of proportionality determined by, e.g.,
$N^{-1}\sum_{\xi_{i}>0}\sigma_{\xi}^{2}(\boldsymbol{x}_{i}|r)=1$. In the
discrete case define
\[
\boldsymbol{S}_{0}=\sum_{\xi_{i}>0}\boldsymbol{f}(\boldsymbol{x}%
_{i})\boldsymbol{f}^{\prime}(\boldsymbol{x}_{i})\xi_{i}\text{ and
}\boldsymbol{S}_{k}=\boldsymbol{S}_{k}\left(  r\right)  =\sum_{\xi_{i}%
>0}\boldsymbol{f}(\boldsymbol{x}_{i})\boldsymbol{f}^{\prime}(\boldsymbol{x}%
_{i})\xi_{i}^{k\left(  1-\frac{r}{2}\right)  }\text{ for }k=1,2.
\]
Note that $\boldsymbol{S}_{0}=\boldsymbol{S}_{1}\left(  0\right)
=\boldsymbol{S}_{2}\left(  1\right)  $. If $\sigma^{2}\left(  \cdot\right)
\in\Sigma_{0}$ then $\mathcal{L}_{\nu=1}\left(  \xi|\sigma\right)  =ch_{\max
}\left(  \boldsymbol{A\boldsymbol{S}}_{1}^{-1}\left(  r\right)  \boldsymbol{S}%
_{2}\left(  r\right)  \boldsymbol{S}_{1}^{-1}\left(  r\right)  \right)  $,
which by Proposition \ref{prop: inequality} \ exceeds $ch_{\max}\left(
\boldsymbol{AA}_{\xi}^{-1}\right)  $; this in turn is minimized by the uniform
design $\xi_{\ast}$, with $\xi_{\ast,i}\equiv1/N$. Thus this approximate
design, which we implement as at (\ref{implement}), is minimax with respect to
bias, over $\Sigma_{0}$. Similarly, in a continuous design space the minimax
bias design is the continuous uniform: $m_{\ast}\left(  \boldsymbol{x}\right)
\equiv1/$\textsc{vol}$\left(  \chi\right)  $.

This discussion has revealed why the minimax designs exhibited in Figures
\ref{fig slr} and \ref{fig quad} become proportional to $\sigma\left(
x\right)  $ as $\nu\rightarrow1$, and for $\nu=1$ are uniform if the
maximization is also carried out over $\Sigma_{0}$. For $\nu<1$, if the form
of $\sigma\left(  \cdot\right)  $ is known then this knowledge can be used to
increase the efficiency of the design, relative to uniformity. In the next
section we show that, if $\sigma\left(  \cdot\right)  $ is unknown but is
allowed to range over $\Sigma_{0}$, then uniform (on their support) designs
are again minimax with respect to \textsc{mse}.

\section{MSE minimizing designs \label{section: minimum mse designs}}

Under (\ref{var funcs}) the maximized, over (\ref{Delta0discrete}), loss
(\ref{max loss|sigma}) is
\[
\mathcal{L}_{\nu}\left(  \xi|r\right)  =\left(  1-\nu\right)  c_{r}%
^{2}tr\left(  \boldsymbol{AS}_{1}^{-1}\left(  r\right)  \boldsymbol{S}%
_{0}\boldsymbol{S}_{1}^{-1}\left(  r\right)  \right)  +\nu ch_{\max}\left(
\boldsymbol{A\boldsymbol{S}}_{1}^{-1}\left(  r\right)  \boldsymbol{S}%
_{2}\left(  r\right)  \boldsymbol{S}_{1}^{-1}\left(  r\right)  \right)  .
\]
Several cases are of interest for fixed $r$. The case $r=0$ corresponds to
homoscedasticity. That for $r=2$ is treated in Kong and Wiens (2014). If $r=1$
(a case which turns out to be least favourable -- see the proof of Lemma
\ref{thm: uniform}), then
\[
\mathcal{L}_{\nu}\left(  \xi|r=1\right)  =\left(  1-\nu\right)  Ntr\left(
\boldsymbol{AS}_{1}^{-1}\left(  1\right)  \boldsymbol{S}_{0}\boldsymbol{S}%
_{1}^{-1}\left(  1\right)  \right)  +\nu ch_{\max}\left(
\boldsymbol{A\boldsymbol{S}}_{1}^{-1}\left(  1\right)  \boldsymbol{S}%
_{0}\boldsymbol{S}_{1}^{-1}\left(  1\right)  \right)  .
\]
In this case the optimal, approximate design $\xi_{\ast}$ is again uniform on
all of $\chi$: $\xi_{\ast,i}\equiv1/N$. And again in the parlance of game
theory, the experimenter's optimal reply to Nature's strategy of placing the
variances proportional to the design weights is to design in such a way that
this variance structure is in fact homoscedastic.

To see that $\xi_{\ast}$ is uniform, note that at this design we have
$\boldsymbol{S}_{0}=\boldsymbol{S}_{2}\left(  1\right)  =\boldsymbol{A}\ $and
$\boldsymbol{S}_{1}\left(  1\right)  =\sqrt{N}\boldsymbol{A}$, so that
$\boldsymbol{S}_{1}^{-1}\left(  1\right)  \boldsymbol{S}_{0}\boldsymbol{S}%
_{1}^{-1}\left(  1\right)  =\left(  N\boldsymbol{A}\right)  ^{-1}$, and it
suffices to note that for any other design $\xi$, by Proposition
\ref{prop: inequality}, $\boldsymbol{S}_{1}^{-1}\left(  1\right)
\boldsymbol{S}_{0}\boldsymbol{S}_{1}^{-1}\left(  1\right)  \succeq
\boldsymbol{A}_{\xi}^{-1}\succeq\left(  N\boldsymbol{A}\right)  ^{-1}$.

Similarly, in the continuous case the uniform design, with density $m_{\ast
}\left(  \boldsymbol{x}\right)  $, minimizes $\mathcal{L}_{\nu}\left(
\xi|r=1\right)  $.

To extend these optimality properties of uniform designs to all of $\Sigma
_{0}$, we first consider the discrete case, and define $\mathcal{L}_{\nu
}\left(  \xi\right)  =\max_{r}\mathcal{L}_{\nu}\left(  \xi|r\right)  $. By the
following lemma, a minimax design is necessarily uniform on its support.

\begin{lemma}
\label{thm: uniform} If $\xi$ is a design with $k$-point support $\left\{
\boldsymbol{x}_{i_{1}},...,\boldsymbol{x}_{i_{k}}\right\}  \subset\chi$
($k\leq N$), placing mass $\xi_{i_{j}}$ at $\boldsymbol{x}_{i_{j}}$, and
$\xi_{k}$ is the design placing mass $1/k$ at each point $\boldsymbol{x}%
_{i_{j}}$, then $\boldsymbol{A}_{\xi}=\sum_{j=1}^{k}\boldsymbol{f}%
(\boldsymbol{x}_{i_{j}})\boldsymbol{f}^{\prime}(\boldsymbol{x}_{i_{j}})$ and
\[
\mathcal{L}_{\nu}\left(  \xi\right)  \geq\mathcal{L}_{\nu}\left(  \xi
_{k}\right)  =\left(  1-\nu\right)  Ntr\left(  \boldsymbol{AA}_{\xi}%
^{-1}\right)  +\nu ch_{\max}\left(  \boldsymbol{AA}_{\xi}^{-1}\right)  .
\]

\end{lemma}

By Lemma \ref{thm: uniform} the search for minimax designs reduces to
searching for support points on which the design is to be uniform. Since
$\boldsymbol{A}_{\xi}$ increases, in the sense of positive semidefiniteness,
as $k$ increases, a minimax design $\xi_{\ast}$ necessarily has maximum
support size. Among approximate designs the optimal choice is thus $\xi
_{\ast,i}\equiv1/N,i=1,...,N$. Among exact designs $\xi_{\ast}$ must have
support size $k_{\ast}=\min(n,N)$; the support points $\left\{  \boldsymbol{x}%
_{i_{1}}^{\ast},...,\boldsymbol{x}_{i_{k_{\ast}}}^{\ast}\right\}  $ are those
minimizing%
\begin{equation}
\mathcal{L}_{\nu}\left(  \xi_{\ast}\right)  =\left(  1-\nu\right)  Ntr\left(
\boldsymbol{AA}_{k_{\ast}}^{-1}\right)  +\nu ch_{\max}\left(  \boldsymbol{AA}%
_{k_{\ast}}^{-1}\right)  , \label{compound}%
\end{equation}
with $\boldsymbol{A}_{k_{\ast}}=\sum_{j=1}^{k_{\ast}}\boldsymbol{f}%
(\boldsymbol{x}_{i_{j}}^{\ast})\boldsymbol{f}^{\prime}(\boldsymbol{x}_{i_{j}%
}^{\ast})$. We are then seeking a compound optimal design, for which problems
some general theory has been furnished by Cook and Wong (1994); in our case
there is however the additional restriction to uniformity.%
\begin{figure}
[tb]
\begin{center}
\includegraphics[
trim=0.000000cm 0.497779cm 0.000000cm 0.498582cm,
height=7.139cm,
width=12.1232cm
]%
{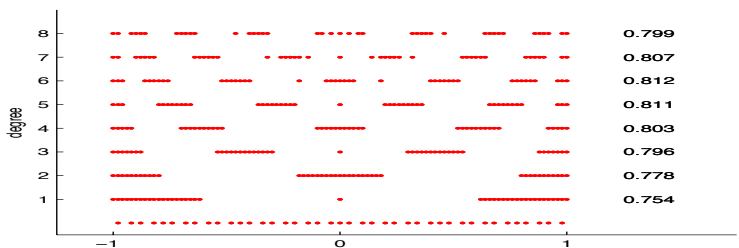}%
\caption{Minimax compound, uniform designs $\xi_{p}$ minimizing $\max
_{\Delta_{0},\Sigma_{0}}$ \textsc{amse} for polynomial regression of degrees
$p=1,...,8$; $n=41$, $N=101$, $\nu=.5$. Bullets indicate design points; bottom
line is the $n$-point implementation of the design $\xi_{\ast,i}\equiv1/N$.
Efficiencies $\mathcal{L}_{\nu}\left(  \xi_{p}\right)  /\mathcal{L}_{\nu
}\left(  \xi_{\ast}\right)  $ are given at the right.}%
\label{fig compound}%
\end{center}
\end{figure}
\medskip

\noindent\textbf{Example 5.1}. In the case of straight line models and
symmetric designs on a symmetric interval, $\boldsymbol{A}_{k_{\ast}%
}=diag\left(  k_{\ast},\sum_{j=1}^{k_{\ast}}\boldsymbol{x}_{i_{j}}^{\ast
2}\right)  $. Both components of $\mathcal{L}_{\nu}\left(  \xi_{\ast}\right)
$ are decreased by progressively including in the support the largest
remaining design points, so as to `increase' $\boldsymbol{A}_{k_{\ast}}$. If
$n$ is odd then $0$ must be in the support; the remaining points -- all points
if $n$ is even -- are the $2\times\min(\left[  n/2\right]  ,\left[
N/2\right]  )$ symmetrically placed design points of largest absolute value.
If $n$ is a multiple of $N$, say $n=mN$, then this design is replicated $m$
times. If $n=mN+t$ for $0<t<N$ then an exact uniform design is not attainable
if $m>0$. A possible implementation is to place $m$ observations at each of
the $N$ points in the design space, and to append to this the $2\left[
t/2\right]  $ symmetrically placed design points of largest absolute value
(and $0$, if $t$ is odd). \medskip

\noindent\textbf{Example 5.2}. We have found an exchange algorithm to be very
effective at constructing compound designs minimizing (\ref{compound}). This
has been carried out for polynomial regression over $\left[  -1,1\right]  $,
with the restriction to symmetric designs. See Figure \ref{fig compound}, in
which some typical cases are displayed and compared with the approximate
design $\xi_{\ast,i}\equiv1/N$, implemented as at (\ref{implement}). The
efficiencies given in the figure have been found to be quite stable over other
choices of $n,N$ and $\nu$.

In a completely analogous manner we find that the continuous uniform design,
with density $m_{\ast}\left(  \boldsymbol{x}\right)  $, minimizes the maximum
of $\mathcal{L}_{\nu}\left(  \xi|r\right)  $ over $\Delta_{0}$ and $\Sigma
_{0}$.

\section{Case study: Robust design in growth charts\label{section: case study}%
}%

\begin{figure}
[tb]
\begin{center}
\includegraphics[
trim=0.667308cm 0.000000cm 0.668507cm 0.000000cm,
height=6.4844cm,
width=16.0573cm
]%
{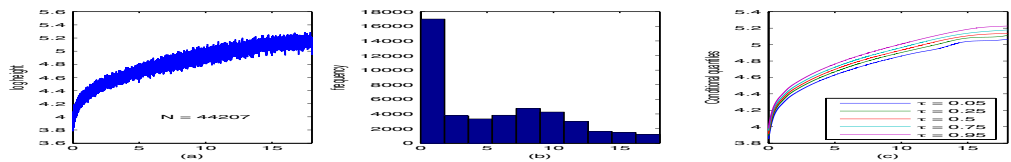}%
\caption{(a) $\log\left(  \text{height}\right)  \ $vs.\ age in full dataset;
(b) frequencies of ages; (c) conditional quantile curves computed from full
dataset.}%
\label{fig: data}%
\end{center}
\end{figure}
%

\begin{figure}
[tb]
\begin{center}
\includegraphics[
trim=0.000000cm 0.000000cm 0.000000cm 0.667851cm,
height=9.6586cm,
width=12.0661cm
]%
{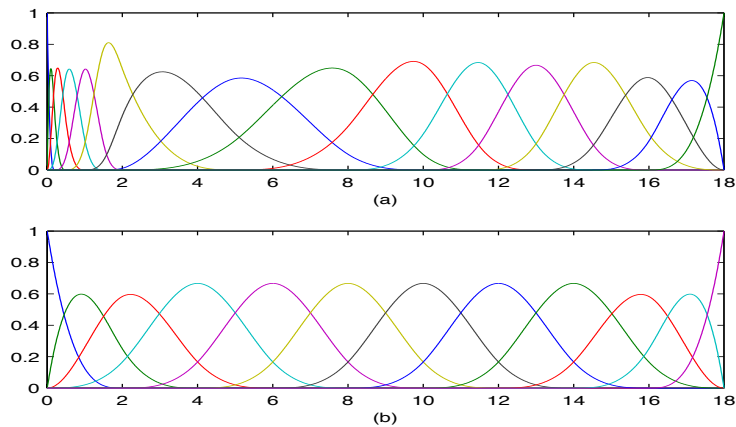}%
\caption{Cubic splines for growth study. (a) `Full' spline basis is of
dimension 16; (b) `Reduced' basis of dimension 12 has fewer and different
internal knots.}%
\label{fig: splines}%
\end{center}
\end{figure}
%

\begin{figure}
[tb]
\begin{center}
\includegraphics[
trim=0.997028cm 0.000000cm 0.998827cm 0.000000cm,
height=6.4844cm,
width=16.154cm
]%
{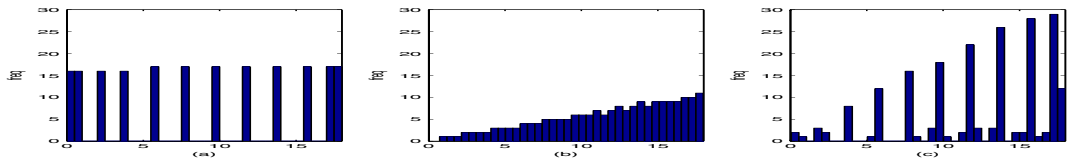}%
\caption{Designs (uniform design not shown) computed for Example 6.1: (a)
Saturated design on 12 points; (b) Minbias design; (c) Minimax design with
$\nu_{0}=.5$.}%
\label{fig: designs1}%
\end{center}
\end{figure}

%

\begin{figure}
[tb]
\begin{center}
\includegraphics[
trim=0.000000cm 0.000000cm 0.000000cm 0.418409cm,
height=9.1775cm,
width=14.467cm
]%
{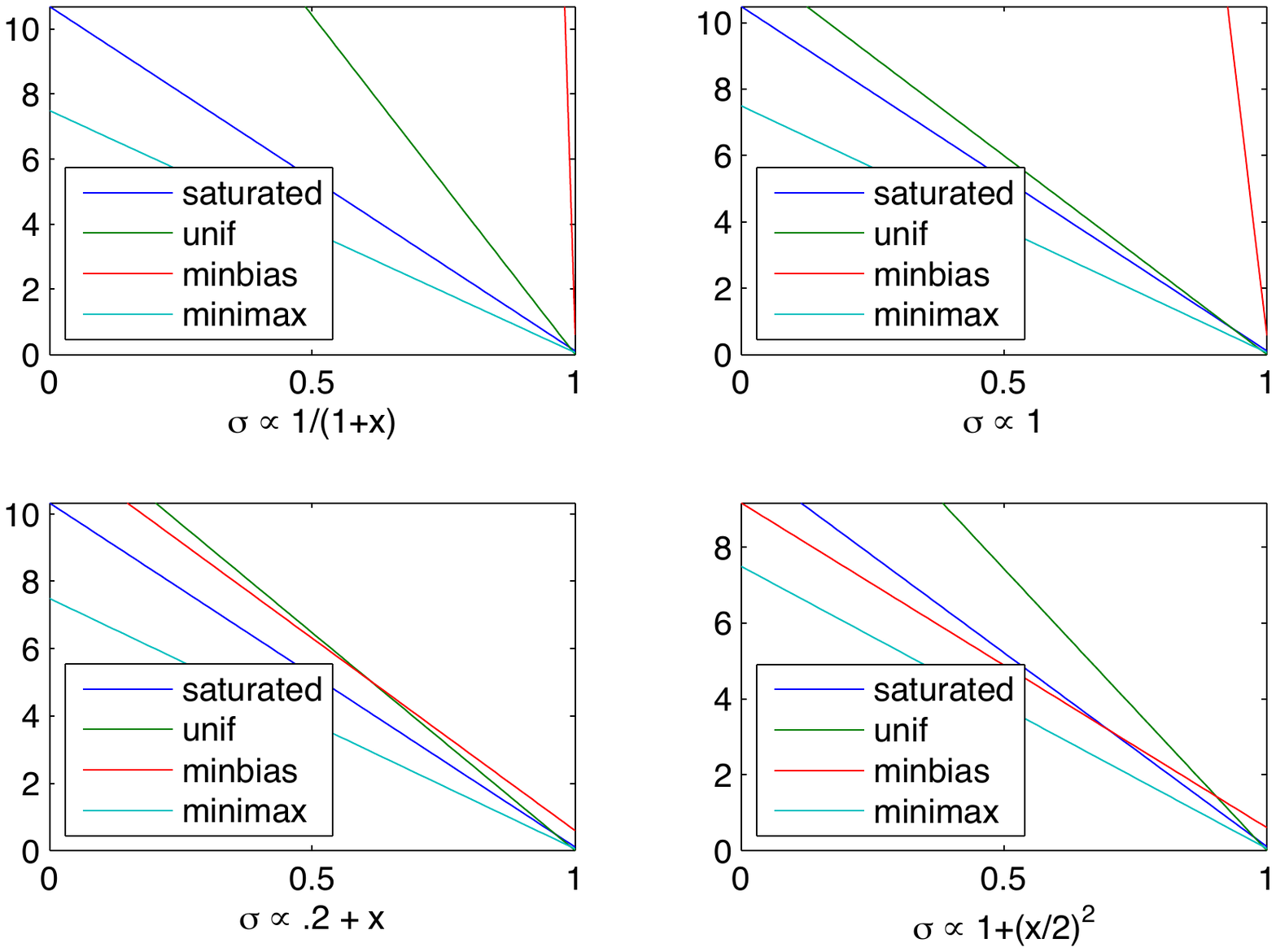}%
\caption{Maximum \textsc{amse} $\mathcal{L}_{\nu}\left(  \xi|\sigma\right)  $
vs.\ $\nu$ for various choices of $\sigma$ and the designs of Example 6.1; the
minimax design was tailored to $\nu_{0}=.5$.}%
\label{fig: amse1}%
\end{center}
\end{figure}

Growth charts, also known as reference centile charts, were first conceived by
Quetelet in the $19^{th}$ century, and are commonly used to screen the
measurements from an individual subject in the context of population values;
to this end they are used by medical practitioners, and others, to monitor
people's growth. A typical growth chart consists of a family of smooth curves
representing a few selected quantiles of the distribution of some physical
measurements -- height, weight, head circumference etc. -- of the reference
population as a function of age. Extreme measurements on the growth chart
suggest that the subject should be studied further, to confirm or to rule out
an unusual underlying physical condition or disease. The conventional method
of constructing growth charts is to get the empirical quantiles of the
measurements at a series of time points, and to then fit a smooth polynomial
curve using the empirical quantiles -- see Hamill, Dridzd, Johnson, Reed,
Roche and Moore (1979). In recent years, a number of different methods have
been developed in the medical statistics literature -- see Wei and He (2006)
for a review.

A recent method proposed by Wei, Pere, Koenker and He (2006) is to estimate a
family of conditional quantile functions by solving nonparametric quantile
regression. In particular, suppose that we want to construct the growth charts
for height. As is common practice in pediatrics, we will take the logarithm of
height ($Y$, in centimeters) as response, and age ($x$, in years), as the
covariate. We consider the nonparametric location/scale model
\[
Y=\mu(x)+\sigma(x)\varepsilon,
\]
where the location function $\mu(x)$ and scale function $\sigma(x)$ satisfy
certain smoothness conditions. Given data $(y_{i},x_{i})$, $i=1,...,n$ the
$\tau^{th}$ quantile curve can be estimated by minimizing $\sum_{i=1}^{n}%
\rho_{\tau}(y_{i}-\mu(x_{i}))$.

For growth charts it is convenient to parameterize the conditional quantile
functions as linear combinations of a few basis functions. Particularly
convenient for this purpose are cubic B-splines. Given a choice of knots for
the B-splines, estimation of the growth charts is a straightforward exercise
in parametric linear regression.

The data -- see Figure \ref{fig: data}, and the detailed description in Pere
(2000) -- were collected retrospectively from health centres and schools in
Finland. To construct the conditional quantile curves in Figure
\ref{fig: data}(c), for ages from birth to $18$ years, we used the entire data
set of size 44207 and the internal knot sequence
\begin{equation}
\{0.2,0.5,1.0,1.5,2.0,5.0,8.0,10.0,11.5,13.0,14.5,16.0\}. \label{bestknots}%
\end{equation}
This sequence was also used by Wei \textit{et al.}\ (2006); see also Kong and
Mizera (2012). Spacing of the internal knots is dictated by the need for more
flexibility during infancy and in the pubertal growth spurt period. Linear
combinations of these functions provide a simple and quite flexible model for
the entire curve over $\left[  0,18\right]  $. Denoting the selected B-splines
by $b_{j}(x)$ $j=1,...,p=16$, we obtain the model (\ref{approx model}) with
$\mu(x)=\boldsymbol{f}^{\prime}(\boldsymbol{x})\boldsymbol{\theta}$ for
$\boldsymbol{f}(\boldsymbol{x})=\left(  b_{1}(x),...,b_{p}(x)\right)
^{\prime}$ and $\boldsymbol{\theta}=\left(  \theta_{1},...,\theta_{p}\right)
^{\prime}$. However, due to uncertainty in the selection of knots and to other
approximations underlying the model, the designer might well seek protection
against departures of the form (\ref{true model}). In this study we will
explore how to sample from the available ages in order to robustly estimate
the growth charts of heights.

In computing and assessing the designs we supposed that the experimenter would
use the internal knot sequence
\begin{equation}
\{2.0,4.0,6.0,8.0,10.0,12.0,14.0,16.0\}; \label{desknots}%
\end{equation}
one measure of design quality is then the accuracy with which the quantile
curves in Figure \ref{fig: data}(c), using the `true' model defined by
(\ref{bestknots}), are recovered from the, much smaller, designed sample
fitted using (\ref{desknots}). \ 

The design space consisted of the $N=1799$ unique values of $x\ $in the
original data set; these span the range $[0,18.0]$ in increments of $.01$ with
only two exceptions. We investigated four types of designs; in all cases
illustrated here we used $n=200$. The first design -- `saturated' -- places
equal weight at each of $p$ points, where $p=12$ is the number of regression
parameters to be estimated in order to fit the reduced cubic spline basis. The
literature provides little guidance on the optimal locations of these points,
but we have followed Kaishev (1989) who studied D-optimal designs for spline
models and conjectured that a `near' optimal design places its mass at the $p$
locations at which the individual splines -- see Figure \ref{fig: splines}%
(b)\ \ -- attain their maxima. Saturated designs enjoy favoured status within
optimal design theory, when there is no doubt that the fitted model is in fact
the correct one. In this current study they turn out to be quite efficient
unless $\nu$ is quite large, i.e.\ loss dominated by bias, in which case both
the uniform and minimax designs, described below, result in predictions with
substantially less bias. As well, the saturated designs are rather poor at
recovering the quantile curves from the data gathered at this small number of locations.

The second design is the uniform, implemented as at (\ref{implement}). This
has been seen to have minimax properties when the maximum is taken over very
broad classes of departures from the nominal model. \ The third -- `minbias'
-- is as described in \S \ref{section: minimum bias designs}, with designs
weights proportional to $\sigma\left(  x\right)  $, again implemented as at
(\ref{implement}). It is not possible to implement such a design very
accurately when $n<N$, and it will be seen that because of this its minimum
bias property is lost. In some cases it does however have attractive behaviour
with respect to the variance component of the \textsc{mse}.

The final design -- `minimax' -- minimizes $\mathcal{L}_{\nu}\left(
\xi|\sigma\right)  $ at (\ref{max loss|sigma}) for a particular variance
function $\sigma^{2}\left(  x\right)  $ chosen from those itemized in the
captions of Figures \ref{fig slr} and \ref{fig quad}. The minimax designs were
obtained using a genetic algorithm similar to that described in Welsh and
Wiens (2013). The algorithm begins by generating a `population' of 40 designs
-- the three designs described above and 37 which are randomly generated. Each
is assigned a `fitness' value, with the designs having the smallest
\textsc{mse} being the `most fit', and a probabilistic mechanism is introduced
by which the most fit members become most likely to be chosen to have
`children'. The children are formed from the parents in a particular way; with
a certain probability they are then subjected to random mutations. In this way
the possible parents in each generation are replaced by their children, thus
forming the next generation of designs. A feature of the algorithm is that a
certain proportion of the members -- the most fit 10\% -- always survive
intact; in essence they become their own children. This ensures that the best
member of\ each generation has \textsc{mse} no larger than that in the
previous generation. In all cases we terminated after 1000 generations without
improvement. \medskip%
\begin{figure}
[tb]
\begin{center}
\includegraphics[
trim=1.351567cm 0.000000cm 0.905244cm 0.632356cm,
height=11.6004cm,
width=15.8091cm
]%
{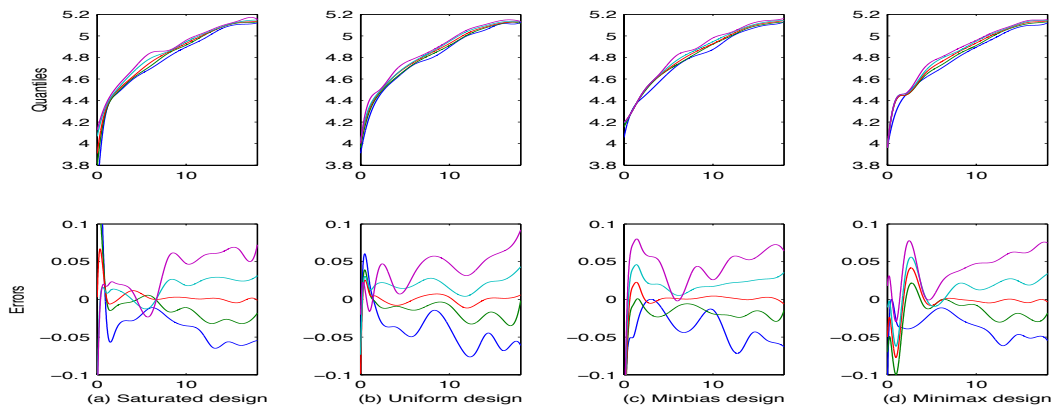}%
\caption{Quantile curves computed in Example 6.1 from the four designs (a) -
(d) and reduced spline basis on knots (\ref{desknots}), and deviations from
those computed using the full dataset and knots (\ref{bestknots}).}%
\label{fig: curves1}%
\end{center}
\end{figure}
%

\begin{figure}
[tb]
\begin{center}
\includegraphics[
height=6.4932cm,
width=12.1232cm
]%
{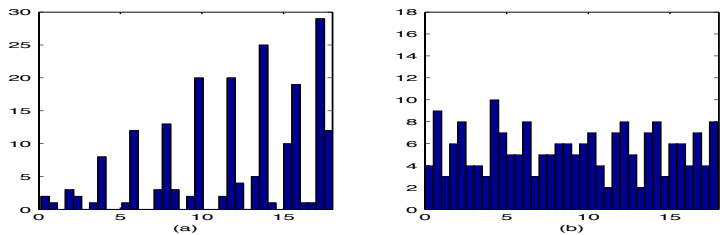}%
\caption{Minimax designs in Examples 6.2 and 6.3; (a) $\nu_{0}=0$; (b)
$\nu_{0}=1$.}%
\label{fig: designs2}%
\end{center}
\end{figure}

\noindent\textbf{Example 6.1}. We computed the four designs, using the
variance function with $\sigma_{0}\left(  x\right)  \propto.2+x$ and, in the
case of the minimax design, a proportion $\nu_{0}=.5$ of the emphasis placed
on bias reduction. See Figure \ref{fig: designs1}. The performance of all
designs against all four of the variance functions is illustrated in Figure
\ref{fig: amse1}, where the maximum \textsc{mse} $\mathcal{L}_{\nu}\left(
\xi|\sigma\right)  $ at (\ref{max loss|sigma}) is plotted against $\nu$. The
efficiency of the minimax design relative to the best of the other three,
which we define in terms of the ratio of the corresponding values of
$\mathcal{L}_{\nu_{0}}\left(  \xi|\sigma_{0}\right)  $, was $1.40$ -- a
substantial gain. We then fit quantile curves, for $\tau=.05,.25,.5,.75,.95$,
to the full data set (Figure \ref{fig: data}(c)) and after each design. See
Figure \ref{fig: curves1}. For each combination of design and $\tau$, root-mse
values were computed as $rmse=\sqrt{\text{mean}\left(  \hat{Y}_{\text{design}%
}-\hat{Y}_{\text{full}}\right)  ^{2}}$, where $\hat{Y}_{\text{full}}$ and
$\hat{Y}_{\text{design}}$ refer to predicted values using the full data set or
those obtained from the designs. This required simulating data, which we did
as follows. To get data at design point $x$ we sampled from a Normal
distribution, with mean given by the value, at $x$, of the `$\tau=.5$' curve
in Figure \ref{fig: data}(c) and variance $\sigma_{Y}^{2}\left(  x\right)  $
estimated from the $Y$-values, at $x$, in the original data. This process was
carried out 100 times; the $rmse$ values given in Table 1 are the averages of
those so obtained, followed by the standard errors in parentheses. The growth
and error curves are based on one representative sample. The uniform and
minimax designs yielded samples from which the quantile curves were recovered
quite accurately; the saturated and minbias designs were generally less
successful. In examples not reported here we found however that for
substantially larger values of $n$ -- for instance $n=1000$ -- the minbias
design performed as well as the others in this regard. \medskip

\noindent\textbf{Example 6.2}. We next took $\nu_{0}=0$ -- all emphasis on
variance reduction -- but otherwise retained the features of Example 6.1. The
saturated, uniform and minbias designs, whose construction does not depend on
$\nu$, were thus as in Example 6.1; the minimax design is in Figure
\ref{fig: designs2}(a) and again enjoyed a relative efficiency of $1.40$ over
the best of the others. The plots of the quantile curves -- not shown -- tell
much the same story as those for Example 6.1. \medskip

\noindent\ \textbf{Example 6.3}. We then took $\nu_{0}=1$ -- all emphasis on
bias reduction -- and obtained the minimax design in Figure
\ref{fig: designs2}(b), with a relative efficiency of $1.62$. See Table 3,
where we give the values of $\mathcal{L}_{\nu}\left(  \xi|\sigma_{0}\right)  $
for all six designs discussed in Examples 5 - 7, at $\nu=0,.5,1$. \medskip

\noindent\textbf{Example 6.4}. As a final example we reran Example 6.1, but
using $\sigma_{0}\left(  x\right)  \propto1/\left(  1+x\right)  $. The minimax
design had a relative efficiency of 1.17 against the best -- the minbias
design -- of the other three; the efficiency was much greater against the
uniform and saturated designs. See Table 2. \medskip%
\begin{figure}
[tb]
\begin{center}
\includegraphics[
height=6.4932cm,
width=12.1232cm
]%
{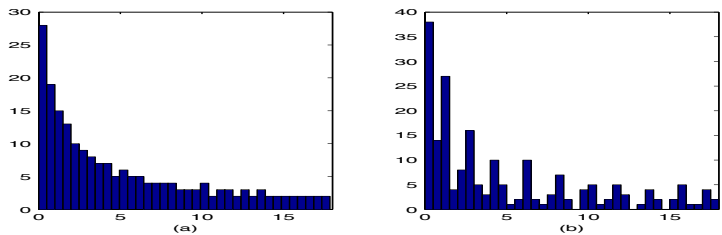}%
\caption{(a) Minbias design and (b) minimax ($\nu_{0}=.5$) design for Example
6.4; both for $\sigma_{0}\left(  x\right)  \propto1/\left(  1+x\right)  $. }%
\label{fig: designs3}%
\end{center}
\end{figure}
%

\begin{table}[tb] \centering
\begin{tabular}
[c]{ccccc}%
\multicolumn{5}{c}{Table 1. Root-mean squared errors (standard errors)}\\
\multicolumn{5}{c}{for the designs in Example 6.1}\\\hline
& \multicolumn{4}{c}{Design}\\\cline{2-5}
& Saturated & Uniform & Minbias & Minimax\\\hline
$\tau=.05$ & \multicolumn{1}{r}{.061 (.0009)} & \multicolumn{1}{r}{.044
(.0003)} & \multicolumn{1}{r}{.060 (.0016)} & \multicolumn{1}{r}{.052
(.0008)}\\
$\tau=.25$ & \multicolumn{1}{r}{.028 (.0005)} & \multicolumn{1}{r}{.020
(.0002)} & \multicolumn{1}{r}{.035 (.0014)} & \multicolumn{1}{r}{.033
(.0012)}\\
$\tau=.50$ & \multicolumn{1}{r}{.012 (.0004)} & \multicolumn{1}{r}{.009
(.0002)} & \multicolumn{1}{r}{.026 (.0015)} & \multicolumn{1}{r}{.026
(.0014)}\\
$\tau=.75$ & \multicolumn{1}{r}{.022 (.0002)} & \multicolumn{1}{r}{.021
(.0002)} & \multicolumn{1}{r}{.031 (.0011)} & \multicolumn{1}{r}{.038
(.0015)}\\
$\tau=.95$ & \multicolumn{1}{r}{.053 (.0007)} & \multicolumn{1}{r}{.046
(.0003)} & \multicolumn{1}{r}{.054 (.0009)} & \multicolumn{1}{r}{.054
(.0010)}\\\hline\hline
\end{tabular}%
\end{table}%
%

\begin{table}[tb] \centering
\begin{tabular}
[c]{ccccccc}%
\multicolumn{7}{c}{Table 2. Maximum \textsc{mse} $\mathcal{L}_{\nu}\left(
\xi|\sigma_{0}\propto.2+x\right)  $ of}\\
\multicolumn{7}{c}{the designs in Examples 6.1 - 6.3}\\\hline
& \multicolumn{6}{c}{Design}\\\cline{2-7}
& Saturated & Uniform & Minbias & \multicolumn{3}{c}{Minimax for:}\\
&  &  &  & $\nu_{0}=0$ & $\nu_{0}=.5$ & $\nu_{0}=1$\\\hline
$\nu=0$ & 10.33 & 12.94 & 12.02 & 7.37 & 7.40 & 12.20\\
$\nu=.5$ & 5.22 & 6.47 & 6.31 & 3.72 & 3.72 & 6.10\\
$\nu=1$ & .111 & .008 & .600 & .068 & .037 & .005\\\hline\hline
\end{tabular}%
\end{table}%
%

\begin{table}[tb] \centering
\begin{tabular}
[c]{ccccc}%
\multicolumn{5}{c}{Table 3. Maximum \textsc{mse} $\mathcal{L}_{\nu}\left(
\xi|\sigma_{0}\propto1/\left(  1+x\right)  \right)  $ of}\\
\multicolumn{5}{c}{the designs in Example 6.4; minimax design uses $\nu
_{0}=.5$}\\\hline
& \multicolumn{4}{c}{Design}\\\cline{2-5}
& Saturated & Uniform & Minbias & Minimax\\
$\nu=0$ & 10.69 & 20.81 & 7.88 & 6.29\\
$\nu=.5$ & 5.40 & 10.41 & 3.95 & 3.15\\
$\nu=1$ & .111 & .006 & .017 & .016\\\hline\hline
\end{tabular}%
\end{table}%

\section{Summary and concluding remarks}

Dette and Trampisch (2012) studied locally optimal quantile regression designs
for nonlinear models, and concluded with a call for future research into the
robustness of designs with respect to the model assumptions. In this article
we have detailed such research, with specific attention to linear models but
with an outline of the modest changes required to address nonlinear models.

Although a number of our methods described here are analytically and
numerically complex, some general guidance is possible. One recurring theme of
this article is that \textit{uniform} designs are often minimax in
sufficiently large classes of the types of departures we consider. It has long
been recognized in problems of design for least squares regression that the
uniform design plays much the same role as does the median in robust
estimation -- highly robust if not terribly efficient -- and our findings seem
to extend this role to quantile regression.

In seeking protection against bias alone, resulting from model
misspecification and a particular variance function, designs with weights
proportional to the root of the variance function turn out to be minimax
against response misspecifications.

Uniform designs and minimum bias designs are easily implemented. The more
complex design strategies illustrated in \S \ref{section: opt var fixed} are
more laborious, but it has been seen that a rough description of the outcomes,
when there are already available non-robust designs which minimize the loss at
the experimenter's assumed model, is that the robust designs can at least be
approximated by taking the replicates prescribed by the non-robust strategies,
and spreading these out into clusters of distinct but nearby design points.

The robust designs obtained here all yield substantial gains in efficiency, as
measured in terms of maximum loss, when compared to their competitors --
enough to warrant some computational complexity in their construction. As is
seen from the plots of the designs -- Figures \ref{fig slr}, \ref{fig quad},
\ref{fig: designs1}, \ref{fig: designs2} and \ref{fig: designs3} in particular
-- a gain in efficiency should be realizable, without a great deal of
computation, by merely following the preceding heuristic of clustering
replicates, and combining this with design weights suggested by the minimum
bias paradigm.

\appendix

\section*{Appendix: Derivations}

\setcounter{equation}{0} \renewcommand{\theequation}{A.\arabic{equation}}

\noindent\textbf{Mathematical developments for
\S \ref{section: fixed var discrete}}. With definitions $\zeta_{i}=\xi
_{i}/\sigma_{i}$, $\boldsymbol{\zeta}=\left(  \zeta_{1},..,\zeta_{N}\right)
^{\prime}$ and%
\[
\gamma_{0}=\frac{1}{N}\sum_{i=1}^{N}x_{i}^{2},\ \kappa_{1}=\sum_{i=1}^{N}%
\zeta_{i},\ \omega_{1}=\sum_{i=1}^{N}\zeta_{i}^{2},\ \gamma_{1}=\sum_{i=1}%
^{N}x_{i}^{2}\sigma_{i}\zeta_{i},\ \kappa_{2}=\sum_{i=1}^{N}x_{i}^{2}\zeta
_{i},\ \omega_{2}=\sum_{i=1}^{N}x_{i}^{2}\zeta_{i}^{2},
\]
(\ref{max loss|sigma}) becomes $\mathcal{L}_{\nu}\left(  \xi\right)  =\left(
1-\nu\right)  \left\{  \frac{1}{\kappa_{1}^{2}}+\frac{\gamma_{0}\gamma_{1}%
}{\kappa_{2}^{2}}\right\}  +\nu\max\left\{  \frac{\omega_{1}}{\kappa_{1}^{2}%
},\frac{\gamma_{0}\omega_{2}}{\kappa_{2}^{2}}\right\}  $. We shall restrict to
variance functions for which we can verify that, evaluated at $\left\{
\xi_{\ast,i}\right\}  _{i=1}^{N}$,
\begin{equation}
\frac{\omega_{1}}{\kappa_{1}^{2}}\geq\frac{\gamma_{0}\omega_{2}}{\kappa
_{2}^{2}}. \label{claim}%
\end{equation}
We thus minimize $\left(  1-\nu\right)  \left\{  \frac{1}{\kappa_{1}^{2}%
}+\frac{\gamma_{0}\gamma_{1}}{\kappa_{2}^{2}}\right\}  +\nu\frac{\omega_{1}%
}{\kappa_{1}^{2}}$, first with $\gamma_{1},\kappa_{1}$ and $\kappa_{2}$ fixed;
we then minimize over these values. For this we minimize $\omega_{1}$, subject
to
\begin{equation}
\text{ }\text{(i) }\sum_{i=1}^{N}x_{i}^{2}\sigma_{i}\zeta_{i}=\gamma
_{1}\text{, (ii) }\sum_{i=1}^{N}\zeta_{i}=\kappa_{1}\text{, (iii) }\sum
_{i=1}^{N}x_{i}^{2}\zeta_{i}=\kappa_{2}\text{, (iv) }\sum_{i=1}^{N}\sigma
_{i}\zeta_{i}=1. \label{constraints}%
\end{equation}
It is sufficient that $\boldsymbol{\zeta}$$\succeq\mathbf{0}$ (i.e., all
elements non-negative) minimize the convex function%
\[
\Phi\left(  \boldsymbol{\zeta},\boldsymbol{\lambda}\right)  =\sum_{i=1}%
^{N}\left[  \zeta_{i}^{2}-2a\left\{  \left(  1+\lambda_{1}x_{i}^{2}\right)
+\sigma_{i}\left(  \lambda_{2}+\lambda_{3}x_{i}^{2}\right)  \right\}
\zeta_{i}\right]  ,
\]
with the multipliers $a\left(  1,\lambda_{1},\lambda_{2},\lambda_{3}\right)
^{\prime}$, pre-arranged in this convenient manner, chosen to satisfy the side
conditions. Since $\Phi$ is a sum of univariate, convex functions it is
minimized over $\boldsymbol{\zeta}$$\succeq\mathbf{0}$ at the pointwise
positive part $\boldsymbol{\zeta}_{0}^{+}\overset{def}{=}\left(  \zeta
_{01}^{+},...,\zeta_{0N}^{+}\right)  ^{\prime}$, where $\boldsymbol{\zeta}%
$$_{0}$ is the stationary point of $\Phi$ and $\zeta_{0i}^{+}=\max\left(
\zeta_{0i},0\right)  $. The calculations yield%
\begin{equation}
\zeta_{\ast i}=\zeta_{\ast i}\left(  \boldsymbol{\lambda}\right)
=\frac{\left\{  \left(  1+\lambda_{1}x_{i}^{2}\right)  +\sigma_{i}\left(
\lambda_{2}+\lambda_{3}x_{i}^{2}\right)  \right\}  ^{+}}{\sum_{i=1}^{N}%
\sigma_{i}\left\{  \left(  1+\lambda_{1}x_{i}^{2}\right)  +\sigma_{i}\left(
\lambda_{2}+\lambda_{3}x_{i}^{2}\right)  \right\}  ^{+}}, \label{r_i}%
\end{equation}
with $\boldsymbol{\lambda}=\left(  \lambda_{1},\lambda_{2},\lambda_{3}\right)
^{\prime}$ determined from (i), (ii) and (iii) of (\ref{constraints}).

We may now minimize over $\boldsymbol{\lambda}$ rather than over $\left(
\gamma_{1},\kappa_{1},\kappa_{2}\right)  $, so that the numerical problem is
to minimize
\[
L\left(  \boldsymbol{\lambda}\right)  =\left(  1-\nu\right)  \left\{  \frac
{1}{\kappa_{1}^{2}}+\frac{\gamma_{0}\gamma_{1}}{\kappa_{2}^{2}}\right\}
+\frac{\nu}{\kappa_{1}^{2}}\sum_{i=1}^{N}\zeta_{i}^{2}\left(
\boldsymbol{\lambda}\right)  ,
\]
with $\zeta_{i}\left(  \boldsymbol{\lambda}\right)  $ defined by (\ref{r_i})
and $\gamma_{1}=\gamma_{1}\left(  \boldsymbol{\lambda}\right)  $, $\kappa
_{1}=\kappa_{1}\left(  \boldsymbol{\lambda}\right)  $, $\kappa_{2}=\kappa
_{2}\left(  \boldsymbol{\lambda}\right)  $ defined by (i), (ii) and (iii) of
(\ref{constraints}). After doing this with a numerical constrained minimizer
we check (\ref{claim}). Then $\xi_{\ast i}=\sigma_{i}\zeta_{\ast i}$.

In Figure \ref{fig slr} we have illustrated only some representative variance
functions for which (\ref{claim}) holds. When it does not, one can minimize
instead $\left(  1-\nu\right)  \left\{  \frac{1}{\kappa_{1}^{2}}+\frac
{\gamma_{0}\gamma_{1}}{\kappa_{2}^{2}}\right\}  +\nu\frac{\gamma_{0}\omega
_{2}}{\kappa_{2}^{2}}$ and then check that, at the optimal design,
$\frac{\gamma_{0}\omega_{2}}{\kappa_{2}^{2}}\geq\frac{\omega_{1}}{\kappa
_{1}^{2}}$. If this also fails, then a more complex method which is however
guaranteed to succeed is that of Daemi and Wiens (2013), used in
\S \ref{section: fixed var continuous}.\hfill$\square\medskip$

\noindent\textbf{Proof of Theorem \ref{thm: max amse discrete}}. By
(\ref{amse}) we are to find%
\[
\max_{\Delta_{0}}\text{\textsc{amse}}=\frac{\tau\left(  1-\tau\right)
}{g_{\varepsilon}^{2}\left(  0\right)  }tr\left(  \boldsymbol{AP}_{1}%
^{-1}\boldsymbol{P}_{0}\boldsymbol{P}_{1}^{-1}\right)  +\max_{\Delta_{0}%
}\left[  \boldsymbol{\mu}_{0}^{\prime}\boldsymbol{P}_{1}^{-1}\boldsymbol{AP}%
_{1}^{-1}\boldsymbol{\mu}_{0}+N^{-1}\sum_{i=1}^{N}\delta_{0}^{2}\left(
\boldsymbol{x}_{i}\right)  \right]  .
\]
We use methods introduced in Fang and Wiens (2000). We first represent the
design by a diagonal matrix $\boldsymbol{D}_{\xi}$ with diagonal elements
$\left\{  \xi_{i}\right\}  $. Define $\boldsymbol{D}_{\sigma}$ to be the
diagonal matrix with diagonal elements $\left\{  \sigma\left(  \boldsymbol{x}%
_{i}\right)  \right\}  $. Let $\boldsymbol{Q}_{1}$ be an $N\times p$ matrix
whose columns form an orthogonal basis for the column space of the matrix
$\boldsymbol{F}$ with rows $\left\{  \boldsymbol{f}^{\prime}\left(
\boldsymbol{x}\right)  \left\vert {}\right.  \boldsymbol{x}\mathbf{\in
}\mathcal{X}\right\}  $ -- recall that this is `Q' in the QR-decomposition of
$\boldsymbol{F}$. Then $\boldsymbol{F}=\boldsymbol{Q}_{1}\boldsymbol{R}$ for a
$p\times p$, nonsingular triangular matrix $\boldsymbol{R}$. Augment
$\boldsymbol{Q}_{1}$ by $\boldsymbol{Q}_{2}:$ $N\times(N-p)$ whose columns
form an orthogonal basis for the orthogonal complement of this space. Then
$[\boldsymbol{Q}_{1}\vdots\boldsymbol{Q}_{2}]$ is an orthogonal matrix and
$\boldsymbol{\delta}_{0}=\left(  \delta_{0}\left(  \boldsymbol{x}_{1}\right)
,...,\delta_{0}\left(  \boldsymbol{x}_{N}\right)  \right)
^{\boldsymbol{\prime}}$ is, by (i) of (\ref{Delta0discrete}), of the form
$\boldsymbol{\delta}_{0}=\eta\boldsymbol{Q}_{2}\boldsymbol{c}$, where
$\left\Vert \boldsymbol{c}\right\Vert \leq1$. In these terms $\boldsymbol{A}%
=N^{-1}\boldsymbol{R}^{\prime}\boldsymbol{R}$ and from (\ref{defn1}) --
(\ref{defn3}),%
\[
\boldsymbol{\mu}_{0}=\eta\boldsymbol{R}^{\prime}\boldsymbol{Q}_{1}^{\prime
}\boldsymbol{D}_{\sigma}^{-1}\boldsymbol{D}_{\xi}\boldsymbol{Q}_{2}%
\boldsymbol{c},\ \boldsymbol{P}_{0}=\boldsymbol{R}^{\prime}\boldsymbol{Q}%
_{1}^{\prime}\boldsymbol{D}_{\xi}\boldsymbol{Q}_{1}\boldsymbol{R}%
,\ \boldsymbol{P}_{1}=\boldsymbol{R}^{\prime}\boldsymbol{Q}_{1}^{\prime
}\boldsymbol{D}_{\sigma}^{-1}\boldsymbol{D}_{\xi}\boldsymbol{Q}_{1}%
\boldsymbol{R}.
\]
Thus
\begin{align*}
&  \max_{\Delta_{0}}\left[  \boldsymbol{\mu}_{0}^{\prime}\boldsymbol{P}%
_{1}^{-1}\boldsymbol{AP}_{1}^{-1}\boldsymbol{\mu}_{0}+N^{-1}\sum_{i=1}%
^{N}\delta_{0}^{2}\left(  \boldsymbol{x}_{i}\right)  \right] \\
&  =\frac{\eta^{2}}{N}\max_{\left\Vert \boldsymbol{c}\right\Vert \leq1}\left[
\left(  \boldsymbol{c}^{\boldsymbol{\prime}}\boldsymbol{Q}_{2}^{\prime
}\boldsymbol{D}_{\xi}\boldsymbol{D}_{\sigma}^{-1}\boldsymbol{Q}_{1}%
\boldsymbol{R}\right)  \left(  \boldsymbol{P}_{1}^{-1}\boldsymbol{R}^{\prime
}\boldsymbol{RP}_{1}^{-1}\right)  \left(  \boldsymbol{R}^{\prime
}\boldsymbol{Q}_{1}^{\prime}\boldsymbol{D}_{\sigma}^{-1}\boldsymbol{D}_{\xi
}\boldsymbol{Q}_{2}\boldsymbol{c}\right)  +\boldsymbol{c}^{\boldsymbol{\prime
}}\boldsymbol{Q}_{2}^{\prime}\boldsymbol{Q}_{2}\boldsymbol{c}\right] \\
&  =\frac{\eta^{2}}{N}ch_{\max}\left[  \boldsymbol{Q}_{2}^{\prime}\left\{
\boldsymbol{D}_{\xi}\boldsymbol{D}_{\sigma}^{-1}\boldsymbol{Q}_{1}%
\boldsymbol{RP}_{1}^{-1}\boldsymbol{R}^{\prime}\boldsymbol{RP}_{1}%
^{-1}\boldsymbol{R}^{\prime}\boldsymbol{Q}_{1}^{\prime}\boldsymbol{D}_{\sigma
}^{-1}\boldsymbol{D}_{\xi}+\mathbf{I}_{N-p}\right\}  \boldsymbol{Q}%
_{2}\right]  .
\end{align*}
Some algebra, followed by a return to the original parameterization, results
in (\ref{max loss|sigma}).\hfill$\square\medskip$

\noindent\textbf{Proof of Theorem \ref{thm: max amse continuous}}. This
parallels the proof of Theorem 1 of Wiens (1992), and can also be obtained by
taking limits, as $N\rightarrow\infty$, in Theorem
\ref{thm: max amse discrete}.\hfill$\square\medskip$

\noindent\textbf{Derivation of (\ref{form of m})}. The, rather lengthy,
calculations for this section are available in Kong and Wiens (2014). As in
\S \ref{section: fixed var discrete}, we consider symmetric designs and
variance functions: $m(x)=m(-x)$ and $\sigma\left(  x\right)  =\sigma\left(
-x\right)  $. In terms of
\[
\mu_{i}=\int_{-1}^{1}x^{i}m(x)dx,\ \kappa_{i}=\int_{-1}^{1}x^{i}\frac
{m(x)}{\sigma\left(  x\right)  }dx,\ \omega_{i}=\int_{-1}^{1}x^{i}\left(
\frac{m(x)}{\sigma\left(  x\right)  }\right)  ^{2}dx,
\]
we define $\pi=2/\left(  \kappa_{4}\kappa_{0}-\kappa_{2}^{2}\right)  ^{2}$,
$\phi_{002}=\pi/\left(  3\kappa_{2}^{2}\right)  $ and
\begin{align*}
\phi_{110}  &  =\pi\left[  \kappa_{4}^{2}-\frac{1}{3}\kappa_{4}\kappa
_{2}\right]  ,\phi_{112}=\pi\left[  \frac{1}{3}\left(  \kappa_{4}\kappa
_{0}+\kappa_{2}^{2}\right)  -2\kappa_{4}\kappa_{2}\right]  ,\phi_{114}%
=\pi\left[  \kappa_{2}^{2}-\frac{1}{3}\kappa_{2}\kappa_{0}\right]  ,\\
\phi_{120}  &  =\pi\left[  \frac{1}{3}\kappa_{2}^{2}-\kappa_{4}\kappa
_{2}\right]  ,\phi_{122}=\pi\left[  \kappa_{4}\kappa_{0}+\kappa_{2}^{2}%
-\frac{2}{3}\kappa_{2}\kappa_{0}\right]  ,\phi_{124}=\pi\left[  \frac{1}%
{3}\kappa_{0}^{2}-\kappa_{2}\kappa_{0}\right]  ,\\
\phi_{210}  &  =\pi\left[  \frac{1}{3}\kappa_{4}^{2}-\frac{1}{5}\kappa
_{4}\kappa_{2}\right]  ,\phi_{212}=\pi\left[  \frac{1}{5}\left(  \kappa
_{4}\kappa_{0}+\kappa_{2}^{2}\right)  -\frac{2}{3}\kappa_{4}\kappa_{2}\right]
,\phi_{214}=\pi\left[  \frac{1}{3}\kappa_{2}^{2}-\frac{1}{5}\kappa_{2}%
\kappa_{0}\right]  ,\\
\phi_{220}  &  =\pi\left[  \frac{1}{5}\kappa_{2}^{2}-\frac{1}{3}\kappa
_{4}\kappa_{2}\right]  ,\phi_{222}=\pi\left[  \frac{1}{3}\left(  \kappa
_{4}\kappa_{0}+\kappa_{2}^{2}\right)  -\frac{2}{5}\kappa_{2}\kappa_{0}\right]
,\phi_{224}=\pi\left[  \frac{1}{5}\kappa_{0}^{2}-\frac{1}{3}\kappa_{2}%
\kappa_{0}\right]  .
\end{align*}
We then calculate that%
\[
tr\left(  \mathbf{A}\boldsymbol{T}_{0}\right)  \overset{def}{=}\rho_{0}\left(
m\right)  =\left[  \phi_{110}+\phi_{220}\right]  +\left[  \phi_{002}%
+\phi_{112}+\phi_{222}\right]  \mu_{2}+\left[  \phi_{114}+\phi_{224}\right]
\mu_{4},
\]
and that%
\[
\mathbf{A}\boldsymbol{T}_{2}=\left(
\begin{array}
[c]{ccc}%
\phi_{110}\omega_{0}+\phi_{112}\omega_{2}+\phi_{114}\omega_{4} & 0 &
\phi_{120}\omega_{0}+\phi_{122}\omega_{2}+\phi_{124}\omega_{4}\\
0 & \phi_{002}\omega_{2} & 0\\
\phi_{210}\omega_{0}+\phi_{212}\omega_{2}+\phi_{214}\omega_{4} & 0 &
\phi_{220}\omega_{0}+\phi_{222}\omega_{2}+\phi_{224}\omega_{4}%
\end{array}
\right)  ,
\]
whose characteristic roots are $\rho_{1}(m)=\phi_{002}\omega_{2}$ and the two
roots of
\[
\left(
\begin{array}
[c]{cc}%
\phi_{110}\omega_{0}+\phi_{112}\omega_{2}+\phi_{114}\omega_{4} & \phi
_{120}\omega_{0}+\phi_{122}\omega_{2}+\phi_{124}\omega_{4}\\
\phi_{210}\omega_{0}+\phi_{212}\omega_{2}+\phi_{214}\omega_{4} & \phi
_{220}\omega_{0}+\phi_{222}\omega_{2}+\phi_{224}\omega_{4}%
\end{array}
\right)  \overset{def}{=}\left(
\begin{array}
[c]{cc}%
\psi_{11} & \psi_{12}\\
\psi_{21} & \psi_{22}%
\end{array}
\right)  .
\]
Of these two roots, one is uniformly greater than the other, and is
\[
\rho_{2}(m)=\frac{\psi_{11}+\psi_{22}}{2}+\left\{  \left(  \frac{\psi
_{11}-\psi_{22}}{2}\right)  ^{2}+\psi_{12}\psi_{21}\right\}  ^{1/2}.
\]
Thus the loss is $\max\left(  \mathcal{L}_{1}\left(  m\right)  ,\mathcal{L}%
_{2}\left(  m\right)  \right)  $, where $\mathcal{L}_{k}\left(  m\right)
=\left(  1-\nu\right)  \rho_{0}(m)+\nu\rho_{k}(m),\ k=1,2$.

We apply Theorem 1 of Daemi and Wiens (2013), by which we may proceed as
follows. We first find a density $m_{1}$ minimizing $\mathcal{L}_{1}\left(
m\right)  $ in the class of densities for which $\mathcal{L}_{1}\left(
m\right)  =\max\left(  \mathcal{L}_{1}\left(  m\right)  ,\mathcal{L}%
_{2}\left(  m\right)  \right)  $, and a density $m_{2}$ minimizing
$\mathcal{L}_{2}\left(  m\right)  $ in the class of densities for which
$\mathcal{L}_{2}\left(  m\right)  =\max\left(  \mathcal{L}_{1}\left(
m\right)  ,\mathcal{L}_{2}\left(  m\right)  \right)  $. Then the optimal
design $\xi_{\ast}$ has density%
\[
m_{\ast}=\left\{
\begin{array}
[c]{cc}%
m_{1}, & \text{if }\mathcal{L}_{1}\left(  m_{1}\right)  \leq\mathcal{L}%
_{2}\left(  m_{2}\right)  ,\\
m_{2}, & \text{if }\mathcal{L}_{2}\left(  m_{2}\right)  \leq\mathcal{L}%
_{1}\left(  m_{1}\right)  .
\end{array}
\right.
\]
The two minimizations are first carried out with $\mu_{2},\mu_{4},\kappa
_{0},\kappa_{2},\kappa_{4}$ held fixed, thus fixing all $\phi_{ijk}$ and
$\rho_{0}(m)$. Under these constraints $\mathcal{L}_{1}\left(  m_{1}\right)
\leq\mathcal{L}_{2}\left(  m_{2}\right)  $ iff $\rho_{1}(m_{1})\leq\rho
_{2}(m_{2})$.

With the aid of Lagrange multipliers we find that both $m_{1}$ and $m_{2}$ are
of the form (\ref{form of m}). The ten constants $a_{ij}$ forming $\mathbf{a}$
are chosen to minimize the loss subject to the side conditions, but it is now
numerically simpler to minimize $\mathcal{L}_{\nu}\left(  \xi|\sigma\right)  $
at (\ref{max loss|sigma}) directly over $\mathbf{a}$, subject to $\int
_{-1}^{1}m(x;\mathbf{a})dx=1$.

The density $m(x;\mathbf{a})$ is overparameterized, and when $\sigma\left(
\cdot\right)  $ is nonconstant we take $a_{01}=1$. In the homogeneous case we
take $a_{02}=1$ and also $a_{i1}\equiv0$ and $a_{00}=0$.\hfill$\square
\medskip$

\noindent\textbf{Proof of\ Proposition \ref{prop: inequality}}$\medskip$. We
give the proof of (i); that of (ii)\ is similar. For $i=1,...,N$ define
$\boldsymbol{b}\left(  \boldsymbol{x}_{i}\right)  =\left(  \mathbf{M}_{p}%
^{-1}p\left(  \boldsymbol{x}_{i}\right)  -\mathbf{M}_{\mathbf{1}}^{-1}\right)
\boldsymbol{f}\left(  \boldsymbol{x}_{i}\right)  I\left(  \boldsymbol{x}%
_{i}\in\chi_{0}\right)  $. Then%
\begin{equation}
0\preceq\sum_{i=1}^{N}\boldsymbol{b}\left(  \boldsymbol{x}_{i}\right)
\boldsymbol{b}^{\prime}\left(  \boldsymbol{x}_{i}\right)  =\mathbf{M}_{p}%
^{-1}\mathbf{M}_{p^{2}}\mathbf{M}_{p}^{-1}-\mathbf{M}_{\mathbf{1}}%
^{-1}.\medskip\tag*{$\square$}%
\end{equation}

\noindent

\noindent\textbf{Proof of Lemma \ref{thm: uniform}}. Write%
\[
\mathcal{L}_{\nu}\left(  \xi|r\right)  =\left(  1-\nu\right)  N\frac{tr\left(
\boldsymbol{AS}_{1}^{-1}\left(  r\right)  \boldsymbol{S}_{0}\boldsymbol{S}%
_{1}^{-1}\left(  r\right)  \right)  }{\sum_{\xi_{i}>0}\xi_{i}^{r}}+\nu
ch_{\max}\left(  \boldsymbol{A\boldsymbol{S}}_{1}^{-1}\left(  r\right)
\boldsymbol{S}_{2}\left(  r\right)  \boldsymbol{S}_{1}^{-1}\left(  r\right)
\right)  ,
\]
and note that $\mathcal{L}_{\nu}\left(  \xi_{k}|r\right)  =\left(
1-\nu\right)  Ntr\left(  \boldsymbol{AA}_{\xi}^{-1}\right)  +\nu ch_{\max
}\left(  \boldsymbol{AA}_{\xi}^{-1}\right)  $, independently of $r$. Thus it
suffices to show that for some $r=r_{\xi}$,
\begin{equation}
\mathcal{L}_{\nu}\left(  \xi|r_{\xi}\right)  \geq\left(  1-\nu\right)
Ntr\left(  \boldsymbol{AA}_{\xi}^{-1}\right)  +\nu ch_{\max}\left(
\boldsymbol{AA}_{\xi}^{-1}\right)  . \label{r_xi}%
\end{equation}
In fact $r_{\xi}=1$ serves the purpose. To see this note that by Proposition
\ref{prop: inequality},%
\[
\frac{tr\left(  \boldsymbol{AS}_{1}^{-1}\left(  1\right)  \boldsymbol{S}%
_{0}\boldsymbol{S}_{1}^{-1}\left(  1\right)  \right)  }{\sum_{\xi_{i}>0}%
\xi_{i}}\geq tr\left(  \boldsymbol{AA}_{\xi}^{-1}\right)  ,
\]
and that for \textit{any} $r$, $\boldsymbol{S}_{1}^{-1}\left(  r\right)
\boldsymbol{S}_{2}\left(  r\right)  \boldsymbol{S}_{1}^{-1}\left(  r\right)
\succeq\boldsymbol{A}_{\xi}^{-1}$, so that also $ch_{\max}\left(
\boldsymbol{A\boldsymbol{S}}_{1}^{-1}\left(  r\right)  \boldsymbol{S}%
_{2}\left(  r\right)  \boldsymbol{S}_{1}^{-1}\left(  r\right)  \right)  \geq
ch_{\max}\left(  \boldsymbol{AA}_{\xi}^{-1}\right)  $. This establishes
(\ref{r_xi}) with $r_{\xi}=1$.\hfill$\square$

\section*{Acknowledgements}

This work has been supported by the Natural Sciences and Engineering Research
Council of Canada.

\section*{References}

\begin{description}
\item Behl, P., Claeskes, G, and Dette, H. (2014), \textquotedblleft Focussed
Model Selection in Quantile Regression,\textquotedblright\ \textit{Statistica
Sinica}, 24, 601-624.

\item Biedermann, S., and Dette, H. (2001), \textquotedblleft Optimal Designs
for Testing the Functional Form of a Regression via Nonparametric Estimation
Techniques,\textquotedblright\ \textit{Statistics and Probability Letters},
52, 215-224.

\item Bischoff, W. (2010), \textquotedblleft An Improvement in the Lack-of-Fit
Optimality of the (Absolutely) Continuous Uniform Design in Respect of Exact
Designs,\textquotedblright\ \textit{Proceedings of the 9th International
Workshop in Model-Oriented Design and Analysis (moda9)}, eds. Giovagnoli,
Atkinson, and Torsney, Springer-Verlag, Berlin Heidelberg.

\item Box, G. E .P., and Draper, N. R.\ (1959), \textquotedblleft A Basis for
the Selection of a Response Surface Design,\textquotedblright\ \textit{Journal
of the American Statistical Association}, 54, 622-654.

\item Cook, R. D., and Wong, W. K. (1994), \textquotedblleft On the
Equivalence of Constrained and Compound Optimal Designs,\textquotedblright%
\textit{\ Journal of the American Statistical Association}, 89, 687-692.

\item Daemi, M., and Wiens, D. P. (2013), \textquotedblleft Techniques for the
Construction of Robust Regression Designs,\textquotedblright\ \textit{The
Canadian Journal of Statistics}, 41, 679 - 695.

\item Dette, H., and Trampisch, M. (2012), \textquotedblleft Optimal Designs
for Quantile Regression Methods,\textquotedblright\ \textit{Journal of the
American Statistical Association}, 107, 1140-1151.

\item Fang, K. T., and Wang, Y. (1994), \textit{Number-Theoretic Methods in
Statistics}, Chapman and Hall.

\item Fang, Z., and Wiens, D. P.\ (2000), \textquotedblleft Integer-Valued,
Minimax Robust Designs for Estimation and Extrapolation in Heteroscedastic,
Approximately Linear Models,\textquotedblright\ \textit{Journal of the
American Statistical Association}, 95, 807-818.

\item Hamill, P. V. V., Dridzd, T. A., Johnson, C. L., Reed, R. B., Roche, A.
F. and Moore, W. M. (1979), \textquotedblleft Physical growth: National Center
for Health Statistics percentiles,\textquotedblright\ \textit{American Journal
of Clinical Nutrition}, 32, 607-629.

\item Heo, G., Schmuland, B., and Wiens, D. P.\ (2001), \textquotedblleft
Restricted Minimax Robust Designs for Misspecified Regression
Models,\textquotedblright\ \textit{The Canadian Journal of Statistics}, 29, 117-128.

\item Huber, P. J.\ (1964), \textquotedblleft Robust Estimation of a Location
Parameter,\textquotedblright\ \textit{The Annals of Mathematical Statistics},
35, 73-101.

\item -------- (1975), \textquotedblleft Robustness and
Designs,\textquotedblright\ in: \textit{A Survey of Statistical Design and
Linear Models}, ed.\ J. N. Srivastava, Amsterdam: North Holland, pp.\ 287-303.

\item -------- (1981), \textit{Robust Statistics}, New York: Wiley.

\item Kaishev, V. K. (1989), \textquotedblleft Optimal Experimental Designs
for the B-spline Regression,\textquotedblright\ \textit{Computational
Statistics \& Data Analysis}, 8, 39-47.

\item Knight, K. (1998), \textquotedblleft Limiting Distributions for $l_{1}$
Estimators Under General Conditions,\textquotedblright\ \textit{Annals of
Statistics}, 26, 755-770.

\item Koenker, R., and Bassett, G. (1978), \textquotedblleft Regression
Quantiles,\textquotedblright\ \textit{Econometrica}, 46, 33-50.

\item Koenker, R. (2005), Quantile Regression. \textit{Cambridge University
Press}.

\item Kong, L., and Mizera, I. (2012), \textquotedblleft Quantile Tomography:
Using Quantiles with Multivariate Data,\textquotedblright\ \textit{Statistica
Sinica}, 22, 1589-1610.

\item Kong, L., and Wiens, D. P. (2014), \textquotedblleft Robust Quantile
Regression Designs,\textquotedblright\ \textit{University of Alberta
Department of Mathematical and Statistical Sciences Technical Report S129},
http://www.stat.ualberta.ca/\symbol{126}wiens/home page/pubs/TR S129.pdf.

\item Li, K .C.\ (1984), \textquotedblleft Robust Regression Designs When the
Design Space Consists of Finitely Many Points,\textquotedblright\ \textit{The
Annals of Statistics}, 12, 269-282.

\item Li, P., and Wiens, D. P. (2011), \textquotedblleft Robustness of Design
for Dose-Response Studies,\textquotedblright\ \textit{Journal of the Royal
Statistical Society (Series B)}, 17, 215-238.

\item Ma, Y., and Wei, Y. (2012), \textquotedblleft Analysis on Censored
Quantile Residual Life Model via Spline Smoothing,\textquotedblright%
\ \textit{Statistica Sinica}, 22, 47-68.

\item Maronna, R. A., and Yohai, V. J.\ (1981), \textquotedblleft Asymptotic
Behaviour of General M-Estimates for Regression and Scale With Random
Carriers,\textquotedblright\ \textit{Zeitschrift f\"{u}r
Wahrscheinlichkeitstheorie und Verwandte Gebiete}, 58, 7-20.

\item Mart\'{\i}nez-Silva, I., Roca-Pardi\~{n}as, J., Lustres-P\'{e}rez, V.,
Lorenzo-Arribas, A., and Cadaro-Su\'{a}rez, C. (2013), \textquotedblleft
Flexible Quantile Regression Models: Application to the Study of the Purple
Sea Urchin,\textquotedblright\ \textit{SORT}, 37, 81-94.

\item Pere, A. (2000), \textquotedblleft Comparison of Two Methods of
Transforming Height and Weight to Normality,\textquotedblright\ \textit{Annals
of Human Biology}, 27, 35-45.

\item Pollard, D. (1991), \textquotedblleft Asymptotics for Least Absolute
Deviation Regression Estimators,\textquotedblright\ \textit{Econometric
Theory}, 7, 186-199.

\item Rubia, A., Sanchis-Marco, L. (2013), \textquotedblleft On Downside Risk
Predictability Through Liquidity and Trading Activity: A Dynamic Quantile
Approach,\textquotedblright\ \textit{International Journal of Forecasting,}
29, 202-219.

\item Shi, P., Ye, J., and Zhou, J. (2003), \textquotedblleft Minimax Robust
Designs for Misspecified Regression Models,\textquotedblright\ \textit{The
Canadian Journal of Statistics}, 31, 397-414.

\item Simpson, D. G., Ruppert, D., and Carroll, R. J.\ (1992),
\textquotedblleft On One-Step GM Estimates and Stability of Inferences in
Linear Regression,\textquotedblright\ \textit{Journal of the American
Statistical Association}, 87, 439-450.

\item Wei, Y., and He, X. (2006), \textquotedblleft Discussion Paper:
Conditional Growth Charts,\textquotedblright\ \textit{Annals of Statistics},
34, 2069-2097.

\item Wei, Y., Pere, A., Koenker, R., and He, X. (2006), \textquotedblleft
Quantile Regression Methods for Reference Growth Charts,\textquotedblright%
\ \textit{Statistics in Medicine}, 25, 1369-1382.

\item Welsh, A. H.\ \ and Wiens, D. P. (2013), \textquotedblleft Robust
Model-based Sampling Designs,\textquotedblright\ \textit{Statistics and
Computing}, 23, 689-701.

\item Wiens, D. P. (1991), \textquotedblleft Designs for Approximately Linear
Regression: Two Optimality Properties of Uniform Designs,\textquotedblright%
\ \textit{Statistics and Probability Letters}; 12, 217-221.

\item -------- (1992), \textquotedblleft Minimax Designs for Approximately
Linear Regression,\textquotedblright\ \textit{Journal of Statistical Planning
and Inference,} 31, 353-371.

\item -------- (2000), \textquotedblleft Robust Weights and Designs for Biased
Regression Models: Least Squares and Generalized
M-Estimation,\textquotedblright\ \textit{Journal of Statistical Planning and
Inference}, 83, 395-412.

\item --------, and Wu, E. K. H. (2010), \textquotedblleft A Comparative Study
of Robust Designs for M-Estimated Regression Models,\textquotedblright%
\ \textit{Computational Statistics and Data Analysis}, 54, 1683-1695.

\item Woods, D.C., Lewis, S.M., Eccleston, J. A., and Russell, K. G.\ (2006),
\ \textquotedblleft Designs for Generalized Linear Models with Several
Variables and Model Uncertainty\textquotedblright,\ \textit{Technometrics},
48, 84--292.

\item Xu, X., and Yuen, W. K. (2011), \textquotedblleft Applications and
Implementations of Continuous Robust Designs,\textquotedblright%
\ \textit{Communications in Statistics - Theory and Methods}, 40, 969-988.
\end{description}
\thispagestyle{empty}

\begin{center}
\textbf{{\large University of Alberta}}

\textbf{{\large Department of Mathematical and Statistical Sciences}}

\textbf{{\large Technical Report S129}}

\bigskip{\large {\textbf{ROBUST QUANTILE\ REGRESSION DESIGNS}}}

\bigskip{\textbf{{\large Linglong Kong and Douglas P. Wiens}}}%
\footnote{\noindent{{\footnotesize Department of Mathematical and Statistical
Sciences; University of Alberta, Edmonton, Alberta; Canada T6G 2G1. e-mail:
\texttt{lkong@ualberta.ca, doug.wiens@ualberta.ca}} }}{\textbf{{\large \ }}%
}$\medskip$

\noindent{\today}
\end{center}

\medskip

\noindent{\textbf{Abstract \ }}This technical report contains unpublished
material, relevant to the article `Model-Robust Designs for Quantile
Regression'. \bigskip

\renewcommand{\theequation}{B.\arabic{equation}}

\section{Proof of Theorem \ref{thm: asymptotic normality}}

The `true' parameter $\boldsymbol{\theta}$ is defined by
\begin{equation}
\boldsymbol{0}=\int_{\chi}E_{Y|\boldsymbol{x}}\left[  \psi_{\tau}\left(
Y-\boldsymbol{f}^{\prime}\left(  \boldsymbol{x}\right)  \boldsymbol{\theta
}\right)  \right]  \boldsymbol{f}\left(  \boldsymbol{x}\right)
d\boldsymbol{x.}\label{root1}%
\end{equation}
The estimate is defined by
\begin{equation}
\boldsymbol{\hat{\theta}}=\arg\min_{\boldsymbol{t}}\sum_{i=1}^{n}\rho_{\tau
}\left(  Y_{i}-\boldsymbol{f}^{\prime}\left(  \boldsymbol{x}_{i}\right)
\boldsymbol{t}\right)  ,\label{thetahat}%
\end{equation}
where $\rho_{\tau}\left(  \cdot\right)  $ is the `check' function $\rho_{\tau
}\left(  r\right)  =r\left(  \tau-I\left(  r<0\right)  \right)  $, with
derivative $\psi_{\tau}\left(  r\right)  =\tau-I\left(  r<0\right)  $. Define
the target parameter $\boldsymbol{\theta}$ to be the asymptotic solution to
(\ref{thetahat}), so that
\begin{equation}
\sum_{i=1}^{n}\xi_{n,i}\psi_{\tau}\left(  Y_{i}-\boldsymbol{f}^{\prime}\left(
\boldsymbol{x}_{i}\right)  \boldsymbol{\theta}\right)  \boldsymbol{f}\left(
\boldsymbol{x}_{i}\right)  \overset{pr}{\rightarrow}\boldsymbol{0}%
,\label{target}%
\end{equation}
in agreement with (\ref{root1}). We require the following conditions.

\begin{description}
\item[(A1)] The distribution function $G_{\varepsilon}$ defined on $\left(
-\infty,\infty\right)  $ is twice continuously differentiable. The density
$g_{\varepsilon}$ is everywhere finite, positive and Lipschitz continuous.

\item[(A2)] $\max_{i=1,\dots,n}\frac{1}{\sqrt{n}}\Vert\boldsymbol{f}%
(\boldsymbol{x}_{i})\mathbf{||}\rightarrow0.$

\item[(A3)] There exists a vector $\boldsymbol{\mu}$, and positive definite
matrices $\boldsymbol{\Sigma}_{0}$ and $\boldsymbol{\Sigma}_{1}$, such that,
with $\delta_{n}^{\ast}(\boldsymbol{x})=$ $\delta_{n}(\boldsymbol{x}%
)/\sigma(\boldsymbol{x})$,
\begin{align*}
\boldsymbol{\mu} &  =\lim_{n\rightarrow\infty}\frac{1}{\sqrt{n}}\sum_{i=1}%
^{n}\left(  \tau-G_{\varepsilon}(-\delta_{n}^{\ast}(\boldsymbol{x}%
_{i}))\right)  \boldsymbol{f}(\boldsymbol{x}_{i}),\\
\boldsymbol{\Sigma}_{0} &  =\lim_{n\rightarrow\infty}\frac{1}{n}\sum_{i=1}%
^{n}G_{\varepsilon}(-\delta_{n}^{\ast}(\boldsymbol{x}_{i}))\left(
1-G_{\varepsilon}(-\delta_{n}^{\ast}(\boldsymbol{x}_{i}))\right)
\boldsymbol{f}(\boldsymbol{x}_{i})\boldsymbol{f}^{\prime}(\boldsymbol{x}%
_{i}),\\
\boldsymbol{\Sigma}_{1} &  =\lim_{n\rightarrow\infty}\frac{1}{n}\sum_{i=1}%
^{n}\frac{g_{\varepsilon}(-\delta_{n}^{\ast}(\boldsymbol{x}_{i}))}%
{\sigma(\boldsymbol{x}_{i})}\boldsymbol{f}(\boldsymbol{x}_{i})\boldsymbol{f}%
^{\prime}(\boldsymbol{x}_{i}).
\end{align*}

\end{description}

\noindent Recall the definitions
\begin{subequations}
\label{denn}%
\begin{align}
\boldsymbol{\mu}_{0} &  =\int_{\chi}\text{ }\delta_{0}(\boldsymbol{x})\frac
{1}{\sigma(\boldsymbol{x})}\boldsymbol{f}(\boldsymbol{x})\xi_{\infty}\left(
d\boldsymbol{x}\right)  ,\label{defn1}\\
\boldsymbol{P}_{0} &  =\int_{\chi}\boldsymbol{f}(\boldsymbol{x})\boldsymbol{f}%
^{\prime}(\boldsymbol{x})\xi_{\infty}\left(  d\boldsymbol{x}\right)
,\label{defn2}\\
\boldsymbol{P}_{1} &  =\int_{\chi}\boldsymbol{f}(\boldsymbol{x})\frac
{1}{\sigma(\boldsymbol{x})}\boldsymbol{f}^{\prime}(\boldsymbol{x})\xi_{\infty
}\left(  d\boldsymbol{x}\right)  .\label{defn3}%
\end{align}
Assume that the support of $\xi_{\infty}$ is large enough that $\boldsymbol{P}%
_{0}$ and $\boldsymbol{P}_{1}$ are positive definite. We have:
\end{subequations}
\begin{theorem}
\label{thm: asymptotic normality}Under conditions (A1) -- (A3) the quantile
regression estimate $\boldsymbol{\hat{\theta}}_{n}$ of the parameter
$\boldsymbol{\theta}$ defined by (\ref{target}) is asymptotically normally
distributed:
\begin{equation}
\sqrt{n}\left(  \boldsymbol{\hat{\theta}}_{n}-\boldsymbol{\theta}\right)
\overset{L}{\rightarrow}N\left(  \boldsymbol{P}_{1}^{-1}\boldsymbol{\mu}%
_{0},\frac{\tau\left(  1-\tau\right)  }{g_{\varepsilon}^{2}\left(  0\right)
}\boldsymbol{P}_{1}^{-1}\boldsymbol{P}_{0}\boldsymbol{P}_{1}^{-1}\right)
.\label{normality}%
\end{equation}

\end{theorem}

\noindent\textbf{Proof} Here we write an $n$-point design as $\left\{
\boldsymbol{x}_{1},...,\boldsymbol{x}_{n}\right\}  $, with the $\boldsymbol{x}%
_{i}\in\chi$ not necessarily distinct. We first show that
\begin{equation}
\sqrt{n}(\hat{\boldsymbol{\theta}}_{n}-\boldsymbol{\theta})\overset
{L}{\rightarrow}N(\boldsymbol{\Sigma}_{1}^{-1}\boldsymbol{\mu}%
,\boldsymbol{\Sigma}_{1}^{-1}\boldsymbol{\Sigma}_{0}\boldsymbol{\Sigma}%
_{1}^{-1}).\label{asym. norm.}%
\end{equation}
For this, define $Z_{n}(\boldsymbol{\gamma})=\sum_{i=1}^{n}\left(  \rho_{\tau
}(u_{i}-f^{\prime}(\boldsymbol{x}_{i})\boldsymbol{\gamma}^{\prime}\sqrt
{n}\mathbf{)-}\rho_{\tau}(u_{i})\right)  $, where $u_{i}=Y_{i}-\boldsymbol{f}%
^{\prime}(\boldsymbol{x}_{i})\boldsymbol{\theta}$ \ and $\hat
{\boldsymbol{\gamma}}=\sqrt{n}(\hat{\boldsymbol{\theta}}_{n}%
-\boldsymbol{\theta})$. The function $Z_{n}(\boldsymbol{\gamma})$ is convex
and is minimized at $\hat{\boldsymbol{\gamma}}$. The main idea of the proof
follows Knight (1998). Using Knight'{}s identity%
\[
\rho_{\tau}(u+v)-\rho_{\tau}(u)=-v\psi_{\tau}(u)+\int_{0}^{v}\left(  I(u\leq
s)-I(u\leq0)\right)  ds,
\]
we may write $Z_{n}(\gamma)=Z_{1n}(\gamma)+Z_{2n}(\gamma)$, where%
\begin{align*}
Z_{1n}(\boldsymbol{\gamma}) &  =-\frac{1}{\sqrt{n}}\sum_{i=1}^{n}%
\boldsymbol{f}^{\prime}(\boldsymbol{x}_{i})\gamma\psi_{\tau}(u_{i}),\\
Z_{2n}(\boldsymbol{\gamma}) &  =\sum_{i=1}^{n}\int_{0}^{v_{ni}}\left(
I(u_{i}\leq s)-I(u_{i}\leq0)\right)  ds\overset{def}{=}\sum_{i=1}^{n}%
Z_{2ni}(\boldsymbol{\gamma}),
\end{align*}
and $v_{ni}=\boldsymbol{\gamma}^{\prime}\boldsymbol{f}(\boldsymbol{x}%
_{i})\sqrt{n}$. We note that%
\[
E[Z_{1n}(\boldsymbol{\gamma})]=-\boldsymbol{\gamma}^{\prime}\frac{1}{\sqrt{n}%
}\sum_{i=1}^{n}\boldsymbol{f}(\boldsymbol{x}_{i})E[\psi_{\tau}(u_{i}%
)]=-\boldsymbol{\gamma}^{\prime}\frac{1}{\sqrt{n}}\sum_{i=1}^{n}\left(
\tau-G_{\varepsilon}(-\delta_{n}^{\ast}(\boldsymbol{x}_{i}))\right)
\boldsymbol{f}(\boldsymbol{x}_{i})
\]
and that%
\begin{align*}
\text{\textsc{var}}[Z_{1n}(\boldsymbol{\gamma})] &  =\boldsymbol{\gamma
}^{\prime}\frac{1}{n}\sum_{i=1}^{n}\boldsymbol{f}^{\prime}(\boldsymbol{x}%
_{i})\boldsymbol{f}(\boldsymbol{x}_{i})\text{\textsc{var}}[\psi_{\tau}%
(u_{i})]\boldsymbol{\gamma}\\
&  =\boldsymbol{\gamma}^{\prime}\frac{1}{n}\sum_{i=1}^{n}G_{\varepsilon
}(-\delta_{n}^{\ast}(\boldsymbol{x}_{i}))\left(  1-G_{\varepsilon}(-\delta
_{n}^{\ast}(\boldsymbol{x}_{i}))\right)  \boldsymbol{f}^{\prime}%
(\boldsymbol{x}_{i})\boldsymbol{f}(\boldsymbol{x}_{i})\boldsymbol{\gamma}.
\end{align*}
It follows from the Lindeberg-Feller Central Limit Theorem, using Condition
(A3), that $Z_{1n}(\boldsymbol{\gamma})\overset{L}{\rightarrow}%
-\boldsymbol{\gamma}^{\prime}\boldsymbol{w}~~$where$~~\boldsymbol{w}\sim
N(\boldsymbol{\mu},\boldsymbol{\Sigma}_{0})$. Now centre $Z_{2n}%
(\boldsymbol{\gamma})$:%
\[
Z_{2n}(\boldsymbol{\gamma})=\sum E[Z_{2ni}(\boldsymbol{\gamma})]+\sum\left(
Z_{2ni}(\boldsymbol{\gamma})-E[Z_{2ni}(\boldsymbol{\gamma})]\right)  .
\]
We have%
\begin{align*}
\sum E[Z_{2ni}(\boldsymbol{\gamma})] &  =\sum\int_{0}^{v_{ni}}\left(
G_{\varepsilon}\left(  -\delta_{n}^{\ast}(\boldsymbol{x}_{i})+\frac{s}%
{\sigma(\boldsymbol{x}_{i})}\right)  -G_{\varepsilon}\left(  -\delta_{n}%
^{\ast}(\boldsymbol{x}_{i})\right)  \right)  ds\\
&  =\frac{1}{n}\sum\int_{0}^{\boldsymbol{f}^{\prime}(\boldsymbol{x}%
_{i})\boldsymbol{\gamma}}g_{\varepsilon}(-\delta_{n}^{\ast}(\boldsymbol{x}%
_{i}))\frac{t}{\sigma(\boldsymbol{x}_{i})}dt+o(1)\\
&  =\frac{1}{2n}\sum\frac{g_{\varepsilon}(-\delta_{n}^{\ast}(\boldsymbol{x}%
_{i}))}{\sigma(\boldsymbol{x}_{i})}\boldsymbol{\gamma}^{\prime}\boldsymbol{f}%
(\boldsymbol{x}_{i})\boldsymbol{f}^{\prime}(\boldsymbol{x}_{i}%
)\boldsymbol{\gamma}+o(1)\\
&  \rightarrow\frac{1}{2}\boldsymbol{\gamma}^{\prime}\boldsymbol{\Sigma}%
_{1}\boldsymbol{\gamma}.
\end{align*}
As well, we have the bound%
\begin{align*}
\text{\textsc{var}}[Z_{2n}(\boldsymbol{\gamma})] &  \leq\sum E\left[  \int
_{0}^{v_{ni}}\left(  I(u_{i}\leq s)-I(u_{i}\leq0)\right)  ds\right]  ^{2}\\
&  \leq\sum E\left[  \int_{0}^{v_{ni}}ds\int_{0}^{v_{ni}}\left(  I(u_{i}\leq
s)-I(u_{i}\leq0)\right)  ds\right]  \\
&  =\sum E\left[  \frac{1}{\sqrt{n}}\boldsymbol{f}^{\prime}(\boldsymbol{x}%
_{i})\boldsymbol{\gamma}\int_{0}^{v_{ni}}\left(  I(u_{i}\leq s)-I(u_{i}%
\leq0)\right)  ds\right]  \\
&  \leq\frac{1}{\sqrt{n}}\max|\boldsymbol{f}^{\prime}(\boldsymbol{x}%
_{i})\boldsymbol{\gamma}|E[Z_{2n}(\boldsymbol{\gamma})].
\end{align*}
Condition (A2) implies that \textsc{var}$[Z_{2n}(\boldsymbol{\gamma
})]\rightarrow0$. As a consequence, $\sum\left(  Z_{2ni}(\boldsymbol{\gamma
})-E[Z_{2ni}(\boldsymbol{\gamma})]\right)  \overset{pr}{\rightarrow}0$ and
$Z_{2n}(\boldsymbol{\gamma})\overset{pr}{\rightarrow}\frac{1}{2}%
\boldsymbol{\gamma}^{\prime}\boldsymbol{\Sigma}_{1}\boldsymbol{\gamma}$.
Combining these observations, we have%
\[
Z_{n}(\boldsymbol{\gamma})\overset{L}{\rightarrow}Z_{0}(\boldsymbol{\gamma
})=-\boldsymbol{\gamma}^{\prime}\boldsymbol{w}+\frac{1}{2}\boldsymbol{\gamma
}^{\prime}\boldsymbol{\Sigma}_{1}\boldsymbol{\gamma}.
\]
The convexity of the limiting objective function $Z_{0}(\boldsymbol{\gamma})$
ensures the uniqueness of the minimizer, which is $\boldsymbol{\gamma}%
_{0}=\boldsymbol{\Sigma}_{1}^{-1}\boldsymbol{w}$. Therefore, we have%
\begin{equation}
\sqrt{n}(\hat{\boldsymbol{\theta}}_{n}-\boldsymbol{\theta})=\hat
{\boldsymbol{\gamma}}=\arg\min Z_{n}(\boldsymbol{\gamma})\overset
{L}{\rightarrow}\boldsymbol{\gamma}_{0}=\arg\min Z_{0}(\boldsymbol{\gamma
}).\label{pfeq5}%
\end{equation}
Similar arguments can be found in Pollard (1991) and Knight (1998). From
\eqref{pfeq5} we immediately obtain (\ref{asym. norm.}).

To go from (\ref{asym. norm.}) to (\ref{normality}) requires passing from the
limits in (A3) to (\ref{denn}). The expansion%
\[
\frac{1}{\sqrt{n}}\left(  \tau-G_{\varepsilon}(-\delta_{n}^{\ast
}(\boldsymbol{x}_{i}))\right)  =\frac{1}{n}\sqrt{n}\left(  G_{\varepsilon
}(0)-G_{\varepsilon}(-\delta_{n}^{\ast}(\boldsymbol{x}_{i}))\right)  =\frac
{1}{n}\left(  g_{\varepsilon}(0)\delta_{0}^{\ast}(\boldsymbol{x}%
_{i})+o(1)\right)
\]
yields $\boldsymbol{\mu}=g_{\varepsilon}(0)\boldsymbol{\mu}_{0}$. Here we
require $\lim\frac{1}{n}\sum_{i=1}^{n}\boldsymbol{f}(\boldsymbol{x}_{i})$ to
be bounded; this is implied by the existence of $\boldsymbol{P}_{0}=\lim
\int_{\chi}\boldsymbol{f}(\boldsymbol{x})\boldsymbol{f}^{\prime}%
(\boldsymbol{x})\xi_{n}\left(  d\boldsymbol{x}\right)  $:%
\[
\Vert\frac{1}{n}\sum_{i=1}^{n}\boldsymbol{f}(\boldsymbol{x}_{i})\Vert^{2}%
\leq\frac{1}{n}\sum_{i=1}^{n}\Vert\boldsymbol{f}(\boldsymbol{x}_{i})\Vert
^{2}=\frac{1}{n}\sum_{i=1}^{n}tr[\boldsymbol{f}(\boldsymbol{x}_{i}%
)\boldsymbol{f}^{\prime}(\boldsymbol{x}_{i})]=tr[\frac{1}{n}\sum_{i=1}%
^{n}\boldsymbol{f}(\boldsymbol{x}_{i})\boldsymbol{f}^{\prime}(\boldsymbol{x}%
_{i})]\rightarrow tr\boldsymbol{P}_{0}.
\]
Similarly, the expansion $G_{\varepsilon}(-\delta_{n}^{\ast}(\boldsymbol{x}%
_{i}))=G_{\varepsilon}(0)-O(n^{-1/2})=\tau-O(n^{-1/2})$ gives that%
\[
\boldsymbol{\Sigma}_{0}=\lim\left\{  \tau(1-\tau)\int_{\chi}\boldsymbol{f}%
(\boldsymbol{x})\boldsymbol{f}^{\prime}(\boldsymbol{x})\xi_{n}\left(
d\boldsymbol{x}\right)  +O(n^{-1/2})\right\}  =\tau(1-\tau)\boldsymbol{P}%
_{0}.
\]
Finally, the expansion $g_{\varepsilon}(-\delta_{n}^{\ast}(\boldsymbol{x}%
_{i}))=g_{\varepsilon}(0)+o(n^{-1/2})$ gives%
\begin{equation}
\boldsymbol{\Sigma}_{1}=\lim\frac{1}{n}\sum_{i=1}^{n}g_{\varepsilon}%
(-\delta_{n}^{\ast}(\boldsymbol{x}_{i}))\boldsymbol{f}(\boldsymbol{x}%
_{i})\boldsymbol{f}^{\prime}(\boldsymbol{x}_{i})/\sigma(\boldsymbol{x}%
_{i})=g_{\varepsilon}(0)\boldsymbol{P}_{1}. \tag*{$\square$}%
\end{equation}

\section{Variance functions $\sigma_{\xi}^{2}\left(  \boldsymbol{x}\right)  $
\label{section: var fns sigma(xi)} - additional examples}

We consider classes $\Sigma_{0}=\left\{  \sigma_{\xi}(\cdot|r)|r\in\left(
-\infty,\infty\right)  \right\}  $ of variance functions given by
\begin{subequations}
\label{choices}%
\begin{align}
\sigma_{\xi}(\boldsymbol{x}_{i}|r)  &  =\left\{
\begin{array}
[c]{cc}%
c_{r}\xi_{i}^{r/2}, & \xi_{i}>0,\\
0, & \xi_{i}=0,
\end{array}
\right.  \text{ with }c_{r}=\left(  \frac{\sum_{\xi_{i}>0}\xi_{i}^{r}}%
{N}\right)  ^{-1/2},\label{discretevar}\\
\sigma_{\xi}(\boldsymbol{x}|r)  &  =\left\{
\begin{array}
[c]{cc}%
c_{r}m^{r/2}\left(  \boldsymbol{x}\right)  , & m(\boldsymbol{x})>0,\\
0, & m(\boldsymbol{x})=0,
\end{array}
\right.  \text{ with }c_{r}=\left(  \int_{m(\boldsymbol{x})>0}m^{r}\left(
\boldsymbol{x}\right)  d\boldsymbol{x}\right)  ^{-1/2}, \label{continuousvar}%
\end{align}
in discrete and continuous spaces respectively. When the experimenter seeks
protection against a \textit{fixed} alternative to homoscedasticity,
i.e.\ fixed $r$, some cases of (\ref{choices})\ may be treated in generality.

Under (\ref{discretevar}) the maximized loss $\mathcal{L}_{\nu}\left(
\xi|\sigma\right)  =\left(  1-\nu\right)  tr\left(  \boldsymbol{AT}%
_{0}\right)  +\nu ch_{\max}\left(  \boldsymbol{AT}_{2}\right)  $ is
\end{subequations}
\begin{equation}
\mathcal{L}_{\nu}\left(  \xi|r\right)  =\left(  1-\nu\right)  c_{r}%
^{2}tr\left(  \boldsymbol{AS}_{1}^{-1}\left(  r\right)  \boldsymbol{S}%
_{0}\boldsymbol{S}_{1}^{-1}\left(  r\right)  \right)  +\nu ch_{\max}\left(
\boldsymbol{A\boldsymbol{S}}_{1}^{-1}\left(  r\right)  \boldsymbol{S}%
_{2}\left(  r\right)  \boldsymbol{S}_{1}^{-1}\left(  r\right)  \right)  ,
\label{max loss|r}%
\end{equation}
where
\begin{align*}
\boldsymbol{S}_{0}  &  =\sum_{\xi_{i}>0}\boldsymbol{f}(\boldsymbol{x}%
_{i})\boldsymbol{f}^{\prime}(\boldsymbol{x}_{i})\xi_{i}\text{,}\\
\boldsymbol{S}_{k}  &  =\boldsymbol{S}_{k}\left(  r\right)  =\sum_{\xi_{i}%
>0}\boldsymbol{f}(\boldsymbol{x}_{i})\boldsymbol{f}^{\prime}(\boldsymbol{x}%
_{i})\xi_{i}^{k\left(  1-\frac{r}{2}\right)  }\text{ for }k=1,2.
\end{align*}
Note that $\boldsymbol{S}_{0}=\boldsymbol{S}_{1}\left(  0\right)
=\boldsymbol{S}_{2}\left(  1\right)  $. \medskip\noindent

\subsection{Discrete designs for variance functions (\ref{choices}) with $r$
fixed}

\noindent\textbf{Example 2.1}. If $r=2$ then $\boldsymbol{S}_{1}%
=\boldsymbol{S}_{2}=\boldsymbol{A}_{\xi}$ and
\begin{equation}
\mathcal{L}_{\nu}\left(  \xi|r=2\right)  =\left(  1-\nu\right)  N\frac
{\sum_{i=1}^{N}\xi_{i}\boldsymbol{f}^{\prime}(\boldsymbol{x}_{i}%
)\boldsymbol{A}_{\xi}^{-1}\boldsymbol{A\boldsymbol{A}}_{\xi}^{-1}%
\boldsymbol{f}(\boldsymbol{x}_{i})}{\sum_{i=1}^{N}\xi_{i}^{2}}+\nu ch_{\max
}\left(  \boldsymbol{A\boldsymbol{A}}_{\xi}^{-1}\right)  . \label{lr2}%
\end{equation}
Without some restriction on the class of designs so as to make it compact,
there are sequences $\left\{  \xi_{\beta}\right\}  $ of designs for which
$\mathcal{L}_{\nu}\left(  \xi_{\beta}\right)  $ tends to the minimum value of
(\ref{lr2}) as $\beta\rightarrow0$, but $\xi_{0}$ has one-point support, so
that $\boldsymbol{A}_{\xi_{0}}$ is singular. To see this, define $s_{0}%
=\min_{1\leq i\leq N}\left\{  \boldsymbol{f}^{\prime}(\boldsymbol{x}%
_{i})\boldsymbol{A}^{-1}\boldsymbol{f}(\boldsymbol{x}_{i})\right\}  $. Since
$\boldsymbol{A}_{\xi}^{-1}\succeq\left(  N\boldsymbol{A}\right)  ^{-1}$ and
$\sum_{i=1}^{N}\xi_{i}^{2}\leq1$, we have that $\mathcal{L}_{\nu}\left(
\xi|r=2\right)  \geq\left(  \left(  1-\nu\right)  s_{0}+\nu\right)
/N\overset{def}{=}\mathcal{L}_{\min}$. If $\xi_{\beta}$ places mass $1-\beta$
at an $\boldsymbol{x}_{\ast}$ for which $s_{0}$ is attained, and mass
$\beta/\left(  N-1\right)  $ at every other point $\boldsymbol{x}_{i}$, then
$\boldsymbol{A}_{\xi_{\beta}}=N\boldsymbol{A}$ and so $\mathcal{L}_{\nu
}\left(  \xi_{\beta}\right)  =\mathcal{L}_{\min}+O\left(  \beta\right)  $ as
$\beta\rightarrow0$. This degeneracy can be avoided by, for instance, imposing
a positive lower bound on the non-zero design weights.

\subsection{Continuous designs for variance functions (\ref{continuousvar})
with $r$ fixed \label{section: continuous, fixed r}}

\noindent\textbf{Example 2.1 continued}. If $r=2$ then $\boldsymbol{S}%
_{1}=\boldsymbol{S}_{2}=\boldsymbol{A}_{m}$ and
\[
\mathcal{L}_{\nu}\left(  \xi|r=2\right)  ={(1-\nu)}\frac{\int_{\chi
}\boldsymbol{f}^{\prime}(\boldsymbol{x})\boldsymbol{A}_{m}^{-1}%
\boldsymbol{A\boldsymbol{A}}_{m}^{-1}\boldsymbol{f}(\boldsymbol{x}%
)m(\boldsymbol{x})d\boldsymbol{x}}{\int_{\chi}m^{2}(\boldsymbol{x}%
)d\boldsymbol{x}}+\nu ch_{\max}\left(  \boldsymbol{A\boldsymbol{A}}_{m}%
^{-1}\right)  .
\]
As in the discrete version of this example, a degenerate solution can be
avoided at the cost of imposing superfluous restrictions on the
designs.\bigskip

\noindent\textbf{Example 2.2} $r=1$. The case $r=1$ and $c_{1}=1$ results in
\[
\mathcal{L}_{\nu}\left(  \xi|r=1\right)  =\left(  1-\nu\right)  tr\left(
\boldsymbol{AS}_{1}^{-1}\left(  1\right)  \boldsymbol{S}_{0}\boldsymbol{S}%
_{1}^{-1}\left(  1\right)  \right)  +\nu ch_{\max}\left(
\boldsymbol{A\boldsymbol{S}}_{1}^{-1}\left(  1\right)  \boldsymbol{S}%
_{0}\boldsymbol{S}_{1}^{-1}\left(  1\right)  \right)  .
\]
The optimal design is uniform, with density $m_{\ast}\left(  \boldsymbol{x}%
\right)  \equiv1/$\textsc{vol}$\left(  \chi\right)  $. To prove this we note
that it is sufficient to show that $\boldsymbol{S}_{1}^{-1}\left(  1\right)
\boldsymbol{S}_{0}\boldsymbol{S}_{1}^{-1}\left(  1\right)  \succeq
\boldsymbol{A}^{-1}$. This is established by introducing $\boldsymbol{A}%
_{m}=\int_{m(\boldsymbol{x})>0}\boldsymbol{f}(\boldsymbol{x})\boldsymbol{f}%
^{\prime}(\boldsymbol{x})d\boldsymbol{x}$ and then using Proposition 1 to
obtain $\boldsymbol{S}_{1}^{-1}\left(  1\right)  \boldsymbol{S}_{0}%
\boldsymbol{S}_{1}^{-1}\left(  1\right)  \succeq\boldsymbol{A}_{m}^{-1}%
\succeq\boldsymbol{A}^{-1}=\boldsymbol{A}_{m_{\ast}}^{-1}$.

\section{\noindent Calculations for the construction of continuous minimax
designs for quadratic regression and fixed variance functions \medskip}

We consider symmetric designs and variance functions: $m(x)=m(-x)$ and
$\sigma\left(  x\right)  =\sigma\left(  -x\right)  $. In terms of
\[
\mu_{i}=\int_{-1}^{1}x^{i}m(x)dx,\kappa_{i}=\int_{-1}^{1}x^{i}\frac
{m(x)}{\sigma\left(  x\right)  }dx,\omega_{i}=\int_{-1}^{1}x^{i}\left(
\frac{m(x)}{\sigma\left(  x\right)  }\right)  ^{2}dx
\]
we have that%
\[
\boldsymbol{T}_{0,0}=\left(
\begin{array}
[c]{ccc}%
1 & 0 & \mu_{2}\\
0 & \mu_{2} & 0\\
\mu_{2} & 0 & \mu_{4}%
\end{array}
\right)  ,\boldsymbol{T}_{0,1}=\left(
\begin{array}
[c]{ccc}%
\kappa_{0} & 0 & \kappa_{2}\\
0 & \kappa_{2} & 0\\
\kappa_{2} & 0 & \kappa_{4}%
\end{array}
\right)  ,\boldsymbol{T}_{0,2}=\left(
\begin{array}
[c]{ccc}%
\omega_{0} & 0 & \omega_{2}\\
0 & \omega_{2} & 0\\
\omega_{2} & 0 & \omega_{4}%
\end{array}
\right)  ,
\]
and
\[
\boldsymbol{T}_{0,1}^{-1}=\frac{1}{\left(  \kappa_{4}\kappa_{0}-\kappa_{2}%
^{2}\right)  }\left(
\begin{array}
[c]{ccc}%
\kappa_{4} & 0 & -\kappa_{2}\\
0 & \kappa_{2}^{-1} & 0\\
-\kappa_{2} & 0 & \kappa_{0}%
\end{array}
\right)  ,\ \mathbf{A}=2\left(
\begin{array}
[c]{ccc}%
1 & 0 & \frac{1}{3}\\
0 & \frac{1}{3} & 0\\
\frac{1}{3} & 0 & \frac{1}{5}%
\end{array}
\right)  .
\]
Define $\pi=\pi\left(  m\right)  =2\left(  \kappa_{4}\kappa_{0}-\kappa_{2}%
^{2}\right)  ^{-2}$. Then
\begin{align*}
\boldsymbol{T}_{2}  &  =\boldsymbol{T}_{0,1}^{-1}\boldsymbol{T}_{0,2}%
\boldsymbol{T}_{0,1}^{-1}\\
&  =\frac{\pi}{2}\left(
\begin{array}
[c]{ccc}%
\kappa_{4} & 0 & -\kappa_{2}\\
0 & \kappa_{2}^{-1} & 0\\
-\kappa_{2} & 0 & \kappa_{0}%
\end{array}
\right)  \left(
\begin{array}
[c]{ccc}%
\omega_{0} & 0 & \omega_{2}\\
0 & \omega_{2} & 0\\
\omega_{2} & 0 & \omega_{4}%
\end{array}
\right)  \left(
\begin{array}
[c]{ccc}%
\kappa_{4} & 0 & -\kappa_{2}\\
0 & \kappa_{2}^{-1} & 0\\
-\kappa_{2} & 0 & \kappa_{0}%
\end{array}
\right) \\
&  =\frac{\pi}{2}\left(
\begin{array}
[c]{ccc}%
\kappa_{4}\omega_{0}-\kappa_{2}\omega_{2} & 0 & \kappa_{4}\omega_{2}%
-\kappa_{2}\omega_{4}\\
0 & \kappa_{2}^{-1}\omega_{2} & 0\\
-\kappa_{2}\omega_{0}+\kappa_{0}\omega_{2} & 0 & -\kappa_{2}\omega_{2}%
+\kappa_{0}\omega_{4}%
\end{array}
\right)  \left(
\begin{array}
[c]{ccc}%
\kappa_{4} & 0 & -\kappa_{2}\\
0 & \kappa_{2}^{-1} & 0\\
-\kappa_{2} & 0 & \kappa_{0}%
\end{array}
\right) \\
&  =\frac{\pi}{2}\left(
\begin{array}
[c]{ccc}%
\begin{array}
[c]{c}%
\kappa_{4}\left(  \kappa_{4}\omega_{0}-\kappa_{2}\omega_{2}\right) \\
-\kappa_{2}\left(  \kappa_{4}\omega_{2}-\kappa_{2}\omega_{4}\right)
\end{array}
& 0 &
\begin{array}
[c]{c}%
-\kappa_{2}\left(  \kappa_{4}\omega_{0}-\kappa_{2}\omega_{2}\right) \\
+\kappa_{0}\left(  \kappa_{4}\omega_{2}-\kappa_{2}\omega_{4}\right)
\end{array}
\\
0 & \frac{\omega_{2}}{\kappa_{2}^{2}} & 0\\%
\begin{array}
[c]{c}%
\kappa_{4}\left(  -\kappa_{2}\omega_{0}+\kappa_{0}\omega_{2}\right) \\
-\kappa_{2}\left(  -\kappa_{2}\omega_{2}+\kappa_{0}\omega_{4}\right)
\end{array}
& 0 &
\begin{array}
[c]{c}%
-\kappa_{2}\left(  -\kappa_{2}\omega_{0}+\kappa_{0}\omega_{2}\right) \\
+\kappa_{0}\left(  -\kappa_{2}\omega_{2}+\kappa_{0}\omega_{4}\right)
\end{array}
\end{array}
\right) \\
&  =\frac{\pi}{2}\left(
\begin{array}
[c]{ccc}%
\kappa_{4}^{2}\omega_{0}-2\kappa_{4}\kappa_{2}\omega_{2}+\kappa_{2}^{2}%
\omega_{4} & 0 &
\begin{array}
[c]{c}%
-\kappa_{4}\kappa_{2}\omega_{0}-\kappa_{2}\kappa_{0}\omega_{4}\\
+\left(  \kappa_{4}\kappa_{0}+\kappa_{2}^{2}\right)  \omega_{2}%
\end{array}
\\
0 & \frac{\omega_{2}}{\kappa_{2}^{2}} & 0\\%
\begin{array}
[c]{c}%
-\kappa_{4}\kappa_{2}\omega_{0}-\kappa_{2}\kappa_{0}\omega_{4}\\
+\left(  \kappa_{4}\kappa_{0}+\kappa_{2}^{2}\right)  \omega_{2}%
\end{array}
& 0 & \kappa_{2}^{2}\omega_{0}-2\kappa_{2}\kappa_{0}\omega_{2}+\kappa_{0}%
^{2}\omega_{4}%
\end{array}
\right)  ;
\end{align*}
hence (replacing $\omega_{i}$ by $\mu_{i}$ in the above)
\[
\boldsymbol{T}_{0}=\boldsymbol{T}_{0,1}^{-1}\boldsymbol{T}_{0,0}%
\boldsymbol{T}_{0,1}^{-1}=\frac{\pi}{2}\left(
\begin{array}
[c]{ccc}%
\kappa_{4}^{2}-2\kappa_{4}\kappa_{2}\mu_{2}+\kappa_{2}^{2}\mu_{4} & 0 &
\begin{array}
[c]{c}%
-\kappa_{4}\kappa_{2}-\kappa_{2}\kappa_{0}\mu_{4}\\
+\left(  \kappa_{4}\kappa_{0}+\kappa_{2}^{2}\right)  \mu_{2}%
\end{array}
\\
0 & \frac{\mu_{2}}{\kappa_{2}^{2}} & 0\\%
\begin{array}
[c]{c}%
-\kappa_{4}\kappa_{2}-\kappa_{2}\kappa_{0}\mu_{4}\\
+\left(  \kappa_{4}\kappa_{0}+\kappa_{2}^{2}\right)  \mu_{2}%
\end{array}
& 0 & \kappa_{2}^{2}-2\kappa_{2}\kappa_{0}\mu_{2}+\kappa_{0}^{2}\mu_{4}%
\end{array}
\right)  .
\]
Then
\begin{align*}
tr\left(  \mathbf{A}\boldsymbol{T}_{0}\right)   &  =\pi tr\left(
\begin{array}
[c]{ccc}%
1 & 0 & 1/3\\
0 & 1/3 & 0\\
1/3 & 0 & 1/5
\end{array}
\right)  \left(
\begin{array}
[c]{ccc}%
\kappa_{4}^{2}-2\kappa_{4}\kappa_{2}\mu_{2}+\kappa_{2}^{2}\mu_{4} & 0 &
\begin{array}
[c]{c}%
-\kappa_{4}\kappa_{2}-\kappa_{2}\kappa_{0}\mu_{4}\\
+\left(  \kappa_{4}\kappa_{0}+\kappa_{2}^{2}\right)  \mu_{2}%
\end{array}
\\
0 & \frac{\mu_{2}}{\kappa_{2}^{2}} & 0\\%
\begin{array}
[c]{c}%
-\kappa_{4}\kappa_{2}-\kappa_{2}\kappa_{0}\mu_{4}\\
+\left(  \kappa_{4}\kappa_{0}+\kappa_{2}^{2}\right)  \mu_{2}%
\end{array}
& 0 & \kappa_{2}^{2}-2\kappa_{2}\kappa_{0}\mu_{2}+\kappa_{0}^{2}\mu_{4}%
\end{array}
\right) \\
&  =\pi\left\{
\begin{array}
[c]{c}%
\left[  \kappa_{4}^{2}-2\kappa_{4}\kappa_{2}\mu_{2}+\kappa_{2}^{2}\mu
_{4}\right]  +\frac{1}{3}\left[  -\kappa_{4}\kappa_{2}-\kappa_{2}\kappa_{0}%
\mu_{4}+\left(  \kappa_{4}\kappa_{0}+\kappa_{2}^{2}\right)  \mu_{2}\right] \\
+\frac{1}{3}\frac{\mu_{2}}{\kappa_{2}^{2}}\\
\frac{1}{3}\left[  -\kappa_{4}\kappa_{2}-\kappa_{2}\kappa_{0}\mu_{4}+\left(
\kappa_{4}\kappa_{0}+\kappa_{2}^{2}\right)  \mu_{2}\right]  +\frac{1}%
{5}\left[  \kappa_{2}^{2}-2\kappa_{2}\kappa_{0}\mu_{2}+\kappa_{0}^{2}\mu
_{4}\right]
\end{array}
\right\} \\
&  =\pi\left\{
\begin{array}
[c]{c}%
\left[  \kappa_{4}^{2}-2\kappa_{4}\kappa_{2}\mu_{2}+\kappa_{2}^{2}\mu
_{4}\right]  +\frac{2}{3}\left[  -\kappa_{4}\kappa_{2}-\kappa_{2}\kappa_{0}%
\mu_{4}+\left(  \kappa_{4}\kappa_{0}+\kappa_{2}^{2}\right)  \mu_{2}\right] \\
+\frac{1}{3}\frac{\mu_{2}}{\kappa_{2}^{2}}+\frac{1}{5}\left[  \kappa_{2}%
^{2}-2\kappa_{2}\kappa_{0}\mu_{2}+\kappa_{0}^{2}\mu_{4}\right]
\end{array}
\right\} \\
&  =\pi\left\{
\begin{array}
[c]{c}%
\left[  \kappa_{4}^{2}-\frac{2}{3}\kappa_{4}\kappa_{2}+\frac{1}{5}\kappa
_{2}^{2}\right]  +\left[  \frac{1}{3\kappa_{2}^{2}}-2\kappa_{4}\kappa
_{2}+\frac{2}{3}\left(  \kappa_{4}\kappa_{0}+\kappa_{2}^{2}\right)  -\frac
{2}{5}\kappa_{2}\kappa_{0}\right]  \mu_{2}\\
+\left[  \kappa_{2}^{2}-\frac{2}{3}\kappa_{2}\kappa_{0}+\frac{1}{5}\kappa
_{0}^{2}\right]  \mu_{4}%
\end{array}
\right\} \\
&  \overset{def}{=}\rho_{0}\left(  m\right)  .
\end{align*}
We have that
\begin{align*}
&  \mathbf{A}\boldsymbol{T}_{2}=\pi\left(
\begin{array}
[c]{ccc}%
1 & 0 & \frac{1}{3}\\
0 & \frac{1}{3} & 0\\
\frac{1}{3} & 0 & \frac{1}{5}%
\end{array}
\right)  \left(
\begin{array}
[c]{ccc}%
\kappa_{4}^{2}\omega_{0}-2\kappa_{4}\kappa_{2}\omega_{2}+\kappa_{2}^{2}%
\omega_{4} & 0 &
\begin{array}
[c]{c}%
-\kappa_{4}\kappa_{2}\omega_{0}-\kappa_{2}\kappa_{0}\omega_{4}\\
+\left(  \kappa_{4}\kappa_{0}+\kappa_{2}^{2}\right)  \omega_{2}%
\end{array}
\\
0 & \frac{\omega_{2}}{\kappa_{2}^{2}} & 0\\%
\begin{array}
[c]{c}%
-\kappa_{4}\kappa_{2}\omega_{0}-\kappa_{2}\kappa_{0}\omega_{4}\\
+\left(  \kappa_{4}\kappa_{0}+\kappa_{2}^{2}\right)  \omega_{2}%
\end{array}
& 0 & \kappa_{2}^{2}\omega_{0}-2\kappa_{2}\kappa_{0}\omega_{2}+\kappa_{0}%
^{2}\omega_{4}%
\end{array}
\right) \\
&  =\pi\left(
\begin{array}
[c]{ccc}%
\begin{array}
[c]{c}%
\left[  \kappa_{4}^{2}\omega_{0}-2\kappa_{4}\kappa_{2}\omega_{2}+\kappa
_{2}^{2}\omega_{4}\right] \\
+\frac{1}{3}\left[  -\kappa_{4}\kappa_{2}\omega_{0}-\kappa_{2}\kappa_{0}%
\omega_{4}+\left(  \kappa_{4}\kappa_{0}+\kappa_{2}^{2}\right)  \omega
_{2}\right]
\end{array}
& 0 &
\begin{array}
[c]{c}%
\left[  -\kappa_{4}\kappa_{2}\omega_{0}-\kappa_{2}\kappa_{0}\omega_{4}+\left(
\kappa_{4}\kappa_{0}+\kappa_{2}^{2}\right)  \omega_{2}\right] \\
+\frac{1}{3}\left[  \kappa_{2}^{2}\omega_{0}-2\kappa_{2}\kappa_{0}\omega
_{2}+\kappa_{0}^{2}\omega_{4}\right]
\end{array}
\\
0 & \frac{\omega_{2}}{3\kappa_{2}^{2}} & 0\\%
\begin{array}
[c]{c}%
\frac{1}{3}\left[  \kappa_{4}^{2}\omega_{0}-2\kappa_{4}\kappa_{2}\omega
_{2}+\kappa_{2}^{2}\omega_{4}\right] \\
+\frac{1}{5}\left[  -\kappa_{4}\kappa_{2}\omega_{0}-\kappa_{2}\kappa_{0}%
\omega_{4}+\left(  \kappa_{4}\kappa_{0}+\kappa_{2}^{2}\right)  \omega
_{2}\right]
\end{array}
& 0 &
\begin{array}
[c]{c}%
\frac{1}{3}\left[  -\kappa_{4}\kappa_{2}\omega_{0}-\kappa_{2}\kappa_{0}%
\omega_{4}+\left(  \kappa_{4}\kappa_{0}+\kappa_{2}^{2}\right)  \omega
_{2}\right] \\
+\frac{1}{5}\left[  \kappa_{2}^{2}\omega_{0}-2\kappa_{2}\kappa_{0}\omega
_{2}+\kappa_{0}^{2}\omega_{4}\right]
\end{array}
\end{array}
\right) \\
&  =\pi\left(
\begin{array}
[c]{ccc}%
\begin{array}
[c]{c}%
\left[  \kappa_{4}^{2}-\frac{1}{3}\kappa_{4}\kappa_{2}\right]  \omega_{0}\\
+\left[  \frac{1}{3}\left(  \kappa_{4}\kappa_{0}+\kappa_{2}^{2}\right)
-2\kappa_{4}\kappa_{2}\right]  \omega_{2}\\
+\left[  \kappa_{2}^{2}-\frac{1}{3}\kappa_{2}\kappa_{0}\right]  \omega_{4}%
\end{array}
& 0 &
\begin{array}
[c]{c}%
\left[  \frac{1}{3}\kappa_{2}^{2}-\kappa_{4}\kappa_{2}\right]  \omega_{0}\\
+\left[  \kappa_{4}\kappa_{0}+\kappa_{2}^{2}-\frac{2}{3}\kappa_{2}\kappa
_{0}\right]  \omega_{2}\\
+\left[  \frac{1}{3}\kappa_{0}^{2}-\kappa_{2}\kappa_{0}\right]  \omega_{4}%
\end{array}
\\
0 & \frac{\omega_{2}}{3\kappa_{2}^{2}} & 0\\%
\begin{array}
[c]{c}%
\left[  \frac{1}{3}\kappa_{4}^{2}-\frac{1}{5}\kappa_{4}\kappa_{2}\right]
\omega_{0}\\
+\left[  \frac{1}{5}\left(  \kappa_{4}\kappa_{0}+\kappa_{2}^{2}\right)
-\frac{2}{3}\kappa_{4}\kappa_{2}\right]  \omega_{2}\\
+\left[  \frac{1}{3}\kappa_{2}^{2}-\frac{1}{5}\kappa_{2}\kappa_{0}\right]
\omega_{4}%
\end{array}
& 0 &
\begin{array}
[c]{c}%
\left[  \frac{1}{5}\kappa_{2}^{2}-\frac{1}{3}\kappa_{4}\kappa_{2}\right]
\omega_{0}\\
+\left[  \frac{1}{3}\left(  \kappa_{4}\kappa_{0}+\kappa_{2}^{2}\right)
-\frac{2}{5}\kappa_{2}\kappa_{0}\right]  \omega_{2}\\
+\left[  \frac{1}{5}\kappa_{0}^{2}-\frac{1}{3}\kappa_{2}\kappa_{0}\right]
\omega_{4}%
\end{array}
\end{array}
\right)  ,
\end{align*}
which we represent as%
\[
\mathbf{A}\boldsymbol{T}_{2}=\left(
\begin{array}
[c]{ccc}%
\phi_{110}\omega_{0}+\phi_{112}\omega_{2}+\phi_{114}\omega_{4} & 0 &
\phi_{120}\omega_{0}+\phi_{122}\omega_{2}+\phi_{124}\omega_{4}\\
0 & \phi_{002}\omega_{2} & 0\\
\phi_{210}\omega_{0}+\phi_{212}\omega_{2}+\phi_{214}\omega_{4} & 0 &
\phi_{220}\omega_{0}+\phi_{222}\omega_{2}+\phi_{224}\omega_{4}%
\end{array}
\right)  ,
\]
where $\phi_{002}=\pi/\left(  3\kappa_{2}^{2}\right)  $ and%
\begin{align*}
\phi_{110}  &  =\pi\left[  \kappa_{4}^{2}-\frac{1}{3}\kappa_{4}\kappa
_{2}\right]  ,\phi_{112}=\pi\left[  \frac{1}{3}\left(  \kappa_{4}\kappa
_{0}+\kappa_{2}^{2}\right)  -2\kappa_{4}\kappa_{2}\right]  ,\phi_{114}%
=\pi\left[  \kappa_{2}^{2}-\frac{1}{3}\kappa_{2}\kappa_{0}\right]  ,\\
\phi_{120}  &  =\pi\left[  \frac{1}{3}\kappa_{2}^{2}-\kappa_{4}\kappa
_{2}\right]  ,\phi_{122}=\pi\left[  \kappa_{4}\kappa_{0}+\kappa_{2}^{2}%
-\frac{2}{3}\kappa_{2}\kappa_{0}\right]  ,\phi_{124}=\pi\left[  \frac{1}%
{3}\kappa_{0}^{2}-\kappa_{2}\kappa_{0}\right]  ,\\
\phi_{210}  &  =\pi\left[  \frac{1}{3}\kappa_{4}^{2}-\frac{1}{5}\kappa
_{4}\kappa_{2}\right]  ,\phi_{212}=\pi\left[  \frac{1}{5}\left(  \kappa
_{4}\kappa_{0}+\kappa_{2}^{2}\right)  -\frac{2}{3}\kappa_{4}\kappa_{2}\right]
,\phi_{214}=\pi\left[  \frac{1}{3}\kappa_{2}^{2}-\frac{1}{5}\kappa_{2}%
\kappa_{0}\right]  ,\\
\phi_{220}  &  =\pi\left[  \frac{1}{5}\kappa_{2}^{2}-\frac{1}{3}\kappa
_{4}\kappa_{2}\right]  ,\phi_{222}=\pi\left[  \frac{1}{3}\left(  \kappa
_{4}\kappa_{0}+\kappa_{2}^{2}\right)  -\frac{2}{5}\kappa_{2}\kappa_{0}\right]
,\phi_{224}=\pi\left[  \frac{1}{5}\kappa_{0}^{2}-\frac{1}{3}\kappa_{2}%
\kappa_{0}\right]  .
\end{align*}
Note that in this notation%
\begin{align*}
\rho_{0}\left(  m\right)   &  =\left\{
\begin{array}
[c]{c}%
\pi\left[  \kappa_{4}^{2}-\frac{2}{3}\kappa_{4}\kappa_{2}+\frac{1}{5}%
\kappa_{2}^{2}\right]  +\pi\left[  \frac{1}{3\kappa_{2}^{2}}-2\kappa_{4}%
\kappa_{2}+\frac{2}{3}\left(  \kappa_{4}\kappa_{0}+\kappa_{2}^{2}\right)
-\frac{2}{5}\kappa_{2}\kappa_{0}\right]  \mu_{2}\\
+\pi\left[  \kappa_{2}^{2}-\frac{2}{3}\kappa_{2}\kappa_{0}+\frac{1}{5}%
\kappa_{0}^{2}\right]  \mu_{4}%
\end{array}
\right\} \\
&  =\left[  \phi_{110}+\phi_{220}\right]  +\left[  \phi_{002}+\phi_{112}%
+\phi_{222}\right]  \mu_{2}+\left[  \phi_{114}+\phi_{224}\right]  \mu_{4}.
\end{align*}
The characteristic roots of $\mathbf{A}\boldsymbol{T}_{2}$ are $\rho
_{1}(m)=\phi_{002}\omega_{2}$ and the two characteristic roots of
\[
\left(
\begin{array}
[c]{cc}%
\phi_{110}\omega_{0}+\phi_{112}\omega_{2}+\phi_{114}\omega_{4} & \phi
_{120}\omega_{0}+\phi_{122}\omega_{2}+\phi_{124}\omega_{4}\\
\phi_{210}\omega_{0}+\phi_{212}\omega_{2}+\phi_{214}\omega_{4} & \phi
_{220}\omega_{0}+\phi_{222}\omega_{2}+\phi_{224}\omega_{4}%
\end{array}
\right)  \overset{def}{=}\left(
\begin{array}
[c]{cc}%
\psi_{11} & \psi_{12}\\
\psi_{21} & \psi_{22}%
\end{array}
\right)  .
\]
Of these two roots, one is uniformly greater than the other, and is
\[
\rho_{2}(m)=\frac{\psi_{11}+\psi_{22}}{2}+\left\{  \left(  \frac{\psi
_{11}-\psi_{22}}{2}\right)  ^{2}+\psi_{12}\psi_{21}\right\}  ^{1/2}.
\]
Thus the loss is the greater of
\begin{align*}
\mathcal{L}_{1}\left(  m\right)   &  =\left(  1-\nu\right)  \rho_{0}%
(m)+\nu\rho_{1}(m),\\
\mathcal{L}_{2}\left(  m\right)   &  =\left(  1-\nu\right)  \rho_{0}%
(m)+\nu\rho_{2}(m).
\end{align*}
We apply Theorem 1 of Daemi and Wiens (2013), by which we may proceed as
follows. We first find a density $m_{1}$ minimizing $\mathcal{L}_{1}\left(
m\right)  $ in the class of densities for which $\mathcal{L}_{1}\left(
m\right)  =\max\left(  \mathcal{L}_{1}\left(  m\right)  ,\mathcal{L}%
_{2}\left(  m\right)  \right)  $, and a density $m_{2}$ minimizing
$\mathcal{L}_{2}\left(  m\right)  $ in the class of densities for which
$\mathcal{L}_{2}\left(  m\right)  =\max\left(  \mathcal{L}_{1}\left(
m\right)  ,\mathcal{L}_{2}\left(  m\right)  \right)  $. Then the optimal
design $\xi_{\ast}$ has density%
\[
m_{\ast}=\left\{
\begin{array}
[c]{cc}%
m_{1}, & \text{if }\mathcal{L}_{1}\left(  m_{1}\right)  \leq\mathcal{L}%
_{2}\left(  m_{2}\right)  ,\\
m_{2}, & \text{if }\mathcal{L}_{2}\left(  m_{2}\right)  \leq\mathcal{L}%
_{1}\left(  m_{1}\right)  .
\end{array}
\right.
\]
The two minimizations are first carried out with $\mu_{2},\mu_{4},\kappa
_{0},\kappa_{2},\kappa_{4}$ held fixed, thus fixing all $\phi_{ijk}$ and
$\rho_{0}(m)$. Under these constraints $\mathcal{L}_{1}\left(  m_{1}\right)
\leq\mathcal{L}_{2}\left(  m_{2}\right)  $ iff $\rho_{1}(m_{1})\leq\rho
_{2}(m_{2})$. We first illustrate the calculations for $m_{2}$.

We seek
\begin{align*}
&  m_{2}=\arg\min\rho_{2}(m),\text{ }\text{subject to}\\
&  \int_{-1}^{1}m\left(  x\right)  dx=1,\int_{-1}^{1}x^{2}m\left(  x\right)
dx=\mu_{2},\int_{-1}^{1}x^{4}m\left(  x\right)  dx=\mu_{4},\\
&  \int_{-1}^{1}\frac{m(x)}{\sigma\left(  x\right)  }dx=\kappa_{0},\int
_{-1}^{1}x^{2}\frac{m(x)}{\sigma\left(  x\right)  }dx=\kappa_{2},\int_{-1}%
^{1}x^{4}\frac{m(x)}{\sigma\left(  x\right)  }dx=\kappa_{4},\\
&  \rho_{2}(m)-\rho_{1}(m)-\beta^{2}=0,
\end{align*}
where $\beta$ is a slack variable. For densities $m\left(  x\right)  $, and
with
\[
m_{\left(  t\right)  }\left(  x\right)  =\left(  1-t\right)  m_{1}\left(
x\right)  +tm\left(  x\right)  ,
\]
it is sufficient to find $m_{1}$ for which the Lagrangian
\begin{align*}
\Phi\left(  t;\boldsymbol{\lambda}\right)   &  =\rho_{2}(m_{\left(  t\right)
})-2\int_{-1}^{1}\left\{  \left[  \lambda_{1}+\frac{\lambda_{4}}{\sigma\left(
x\right)  }\right]  +x^{2}\left[  \lambda_{2}+\frac{\lambda_{5}}{\sigma\left(
x\right)  }\right]  +x^{4}\left[  \lambda_{3}+\frac{\lambda_{6}}{\sigma\left(
x\right)  }\right]  \right\}  m_{\left(  t\right)  }dx\\
&  -\lambda_{7}\left(  \rho_{2}(m_{\left(  t\right)  })-\rho_{1}(m_{\left(
t\right)  })\right)
\end{align*}
is minimized at $t=0$ for every $m\left(  \cdot\right)  $ and satisfies the
side conditions. The first order condition is%
\begin{align}
0  &  \leq\Phi^{\prime}\left(  0;\boldsymbol{\lambda}\right)  =\left(
1-\lambda_{7}\right)  \frac{d}{dt}\rho_{2}\left(  m_{\left(  t\right)
}\right)  _{|_{t=0}}+\lambda_{7}\frac{d}{dt}\rho_{1}\left(  m_{\left(
t\right)  }\right)  _{|_{t=0}}\nonumber\\
&  -2\int_{-1}^{1}\left\{  \left[  \lambda_{1}+\frac{\lambda_{4}}%
{\sigma\left(  x\right)  }\right]  +x^{2}\left[  \lambda_{2}+\frac{\lambda
_{5}}{\sigma\left(  x\right)  }\right]  +x^{4}\left[  \lambda_{3}%
+\frac{\lambda_{6}}{\sigma\left(  x\right)  }\right]  \right\}  \left(
m\left(  x\right)  -m_{2}\left(  x\right)  \right)  dx. \label{Phiprime}%
\end{align}
We have that
\[
\frac{d}{dt}\rho_{1}\left(  m_{\left(  t\right)  }\right)  _{|_{t=0}}%
=2\phi_{002}\int_{-1}^{1}\left(  \frac{m_{2}(x)}{\sigma\left(  x\right)
}\right)  \left(  m\left(  x\right)  -m_{2}\left(  x\right)  \right)  dx,
\]
and, with $\rho_{2}(m)$ represented in an obvious manner as $\rho_{2}%
(m)=\psi_{0}\left(  m\right)  +\sqrt{\psi_{1}\left(  m\right)  }$,
\begin{align*}
&  \frac{d}{dt}\rho_{2}\left(  m_{\left(  t\right)  }\right)  _{|_{t=0}}\\
&  =\frac{d}{dt}\psi_{0}\left(  m_{\left(  t\right)  }\right)  _{|_{t=0}%
}+\frac{1}{2\sqrt{\psi_{1}\left(  m_{2}\right)  }}\frac{d}{dt}\psi_{1}\left(
m_{\left(  t\right)  }\right)  _{|_{t=0}}\\
&  =\frac{1}{2}\frac{d}{dt}\left\{  \int_{-1}^{1}\left\{  \left[  \phi
_{110}+\phi_{220}\right]  +\left[  \phi_{112}+\phi_{222}\right]  x^{2}+\left[
\phi_{114}+\phi_{224}\right]  x^{4}\right\}  \left(  \frac{m_{\left(
t\right)  }(x)}{\sigma\left(  x\right)  }\right)  ^{2}dx\right\}  _{|_{t=0}}\\
&  +\frac{1}{2\sqrt{\psi_{1}\left(  m_{2}\right)  }}\left\{
\begin{array}
[c]{c}%
\left[  2\frac{\psi_{11}\left(  m_{2}\right)  -\psi_{22}\left(  m_{2}\right)
}{2}\right]  \cdot\frac{1}{2}\frac{d}{dt}\left[  \psi_{11}\left(  m_{\left(
t\right)  }\right)  -\psi_{22}\left(  m_{\left(  t\right)  }\right)  \right]
\\
+\frac{d}{dt}\psi_{12}\left(  m_{\left(  t\right)  }\right)  \psi_{21}\left(
m_{2}\right)  +\psi_{12}\left(  m_{2}\right)  \frac{d}{dt}\psi_{21}\left(
m_{\left(  t\right)  }\right)
\end{array}
\right\}  _{|_{t=0}},
\end{align*}
which continues as%
\begin{align*}
&  \frac{d}{dt}\rho_{2}\left(  m_{\left(  t\right)  }\right)  _{|_{t=0}}\\
&  =\int_{-1}^{1}\left\{  \left[  \phi_{110}+\phi_{220}\right]  +\left[
\phi_{112}+\phi_{222}\right]  x^{2}+\left[  \phi_{114}+\phi_{224}\right]
x^{4}\right\}  \left(  \frac{m_{2}(x)}{\sigma^{2}\left(  x\right)  }\right)
\left(  m\left(  x\right)  -m_{2}\left(  x\right)  \right)  dx\\
&  +\frac{1}{2\sqrt{\psi_{1}\left(  m_{2}\right)  }}\left\{
\begin{array}
[c]{c}%
\left[  \psi_{11}\left(  m_{2}\right)  -\psi_{22}\left(  m_{2}\right)
\right]  \int_{-1}^{1}\left\{
\begin{array}
[c]{c}%
\left[  \phi_{110}-\phi_{220}\right] \\
+\left[  \phi_{112}-\phi_{222}\right]  x^{2}\\
+\left[  \phi_{114}-\phi_{224}\right]  x^{4}%
\end{array}
\right\}  \left(  \frac{m_{2}(x)}{\sigma^{2}\left(  x\right)  }\right)
\left(  m\left(  x\right)  -m_{2}\left(  x\right)  \right)  dx\\
+\psi_{21}\left(  m_{2}\right)  \int_{-1}^{1}\left\{  \phi_{120}+\phi
_{122}x^{2}+\phi_{124}x^{4}\right\}  \left(  \frac{m_{2}(x)}{\sigma^{2}\left(
x\right)  }\right)  \left(  m\left(  x\right)  -m_{2}\left(  x\right)
\right)  dx\\
+\psi_{12}\left(  m_{2}\right)  \int_{-1}^{1}\left\{  \phi_{210}+\phi
_{212}x^{2}+\phi_{214}x^{4}\right\}  \left(  \frac{m_{2}(x)}{\sigma^{2}\left(
x\right)  }\right)  \left(  m\left(  x\right)  -m_{2}\left(  x\right)
\right)  dx
\end{array}
\right\} \\
&  =\int_{-1}^{1}\left(  K_{0}+K_{2}x^{2}+K_{4}x^{4}\right)  \left(
\frac{m_{2}(x)}{\sigma^{2}\left(  x\right)  }\right)  \left(  m\left(
x\right)  -m_{2}\left(  x\right)  \right)  dx,
\end{align*}
for%
\begin{align*}
K_{0}  &  =\phi_{110}+\phi_{220}+\frac{\left[  \psi_{11}\left(  m_{2}\right)
-\psi_{22}\left(  m_{2}\right)  \right]  \left[  \phi_{110}-\phi_{220}\right]
+\psi_{21}\left(  m_{2}\right)  \phi_{120}+\psi_{12}\left(  m_{2}\right)
\phi_{210}}{2\sqrt{\psi_{1}\left(  m_{2}\right)  }},\\
K_{2}  &  =\phi_{112}+\phi_{222}+\frac{\left[  \psi_{11}\left(  m_{2}\right)
-\psi_{22}\left(  m_{2}\right)  \right]  \left[  \phi_{112}-\phi_{222}\right]
+\psi_{21}\left(  m_{2}\right)  \phi_{122}+\psi_{12}\left(  m_{2}\right)
\phi_{212}}{2\sqrt{\psi_{1}\left(  m_{2}\right)  }},\\
K_{4}  &  =\phi_{114}+\phi_{224}+\frac{\left[  \psi_{11}\left(  m_{2}\right)
-\psi_{22}\left(  m_{2}\right)  \right]  \left[  \phi_{114}-\phi_{224}\right]
+\psi_{21}\left(  m_{2}\right)  \phi_{124}+\psi_{12}\left(  m_{2}\right)
\phi_{214}}{2\sqrt{\psi_{1}\left(  m_{2}\right)  }}.
\end{align*}
Substituting into (\ref{Phiprime}) gives\newline%
\begin{align*}
\Phi^{\prime}\left(  0;\boldsymbol{\lambda}\right)   &  =\left(  1-\lambda
_{7}\right)  \int_{-1}^{1}\left(  K_{0}+K_{2}x^{2}+K_{4}x^{4}\right)  \left(
\frac{m_{2}(x)}{\sigma^{2}\left(  x\right)  }\right)  \left(  m\left(
x\right)  -m_{2}\left(  x\right)  \right)  dx\\
&  +\lambda_{7}\cdot2\phi_{002}\int_{-1}^{1}\left(  \frac{m_{2}(x)}%
{\sigma\left(  x\right)  }\right)  \left(  m\left(  x\right)  -m_{2}\left(
x\right)  \right)  dx\\
&  -2\int_{-1}^{1}\left\{  \left[  \lambda_{1}+\frac{\lambda_{4}}%
{\sigma\left(  x\right)  }\right]  +x^{2}\left[  \lambda_{2}+\frac{\lambda
_{5}}{\sigma\left(  x\right)  }\right]  +x^{4}\left[  \lambda_{3}%
+\frac{\lambda_{6}}{\sigma\left(  x\right)  }\right]  \right\}  \left(
m\left(  x\right)  -m_{2}\left(  x\right)  \right)  dx\\
&  =\int_{-1}^{1}\left\{
\begin{array}
[c]{c}%
\left\{  \left[  \left(  1-\lambda_{7}\right)  \left(  \frac{K_{0}+K_{2}%
x^{2}+K_{4}x^{4}}{\sigma^{2}\left(  x\right)  }\right)  +\frac{2\lambda
_{7}\phi_{002}}{\sigma\left(  x\right)  }\right]  m_{2}(x)\right\} \\
-2\left\{  \left[  \lambda_{1}+\frac{\lambda_{4}}{\sigma\left(  x\right)
}\right]  +x^{2}\left[  \lambda_{2}+\frac{\lambda_{5}}{\sigma\left(  x\right)
}\right]  +x^{4}\left[  \lambda_{3}+\frac{\lambda_{6}}{\sigma\left(  x\right)
}\right]  \right\}
\end{array}
\right\}  \left(  m\left(  x\right)  -m_{2}\left(  x\right)  \right)  dx,
\end{align*}
entailing%
\[
m_{2}(x)=\left(  \frac{2\left\{  \left[  \lambda_{1}+\frac{\lambda_{4}}%
{\sigma\left(  x\right)  }\right]  +x^{2}\left[  \lambda_{2}+\frac{\lambda
_{5}}{\sigma\left(  x\right)  }\right]  +x^{4}\left[  \lambda_{3}%
+\frac{\lambda_{6}}{\sigma\left(  x\right)  }\right]  \right\}  }{\left[
\left(  1-\lambda_{7}\right)  \left(  \frac{K_{0}+K_{2}x^{2}+K_{4}x^{4}%
}{\sigma^{2}\left(  x\right)  }\right)  +\frac{2\lambda_{7}\phi_{002}}%
{\sigma\left(  x\right)  }\right]  }\right)  ^{+}.
\]

The derivation of $m_{1}$ is very similar. We seek
\begin{align*}
&  m_{1}=\arg\min\rho_{1}(m),\text{ }\text{subject to}\\
&  \int_{-1}^{1}m\left(  x\right)  dx=1,\int_{-1}^{1}x^{2}m\left(  x\right)
dx=\mu_{2},\int_{-1}^{1}x^{4}m\left(  x\right)  dx=\mu_{4},\\
&  \int_{-1}^{1}\frac{m(x)}{\sigma\left(  x\right)  }dx=\kappa_{0},\int
_{-1}^{1}x^{2}\frac{m(x)}{\sigma\left(  x\right)  }dx=\kappa_{2},\int_{-1}%
^{1}x^{4}\frac{m(x)}{\sigma\left(  x\right)  }dx=\kappa_{4},\\
&  \rho_{1}(m)-\rho_{2}(m)-\beta^{2}=0,
\end{align*}
where $\beta$ is a slack variable. For densities $m\left(  x\right)  $, and
with
\[
m_{\left(  t\right)  }\left(  x\right)  =\left(  1-t\right)  m_{2}\left(
x\right)  +tm\left(  x\right)  ,
\]
it is sufficient to find $m_{2}$ for which the Lagrangian
\begin{align*}
\Phi\left(  t;\boldsymbol{\lambda}\right)   &  =\rho_{1}(m_{\left(  t\right)
})-2\int_{-1}^{1}\left\{  \left[  \lambda_{1}+\frac{\lambda_{4}}{\sigma\left(
x\right)  }\right]  +x^{2}\left[  \lambda_{2}+\frac{\lambda_{5}}{\sigma\left(
x\right)  }\right]  +x^{4}\left[  \lambda_{3}+\frac{\lambda_{6}}{\sigma\left(
x\right)  }\right]  \right\}  m_{\left(  t\right)  }dx\\
&  -\lambda_{7}\left(  \rho_{1}(m_{\left(  t\right)  })-\rho_{2}(m_{\left(
t\right)  })\right)
\end{align*}
is minimized at $t=0$ for every $m\left(  \cdot\right)  $ and satisfies the
side conditions. This leads to the same first order condition as
(\ref{Phiprime}), except that $\lambda_{7}$ is replaced by $1-\lambda_{7}$;
this in turn leads to
\[
m_{1}(x)=\left(  \frac{2\left\{  \left[  \lambda_{1}+\frac{\lambda_{4}}%
{\sigma\left(  x\right)  }\right]  +x^{2}\left[  \lambda_{2}+\frac{\lambda
_{5}}{\sigma\left(  x\right)  }\right]  +x^{4}\left[  \lambda_{3}%
+\frac{\lambda_{6}}{\sigma\left(  x\right)  }\right]  \right\}  }{\left[
\lambda_{7}\left(  \frac{K_{0}+K_{2}x^{2}+K_{4}x^{4}}{\sigma^{2}\left(
x\right)  }\right)  +\frac{2\left(  1-\lambda_{7}\right)  \phi_{002}}%
{\sigma\left(  x\right)  }\right]  }\right)  ^{+}.
\]

In either case, the minimizing design density is of the form%
\begin{equation}
m(x;\mathbf{a})=\left(  \frac{q_{1}\left(  x\right)  +\frac{q_{2}\left(
x\right)  }{\sigma\left(  x\right)  }}{\frac{a_{00}}{\sigma\left(  x\right)
}+\frac{q_{3}\left(  x\right)  }{\sigma^{2}\left(  x\right)  }}\right)  ^{+},
\label{form}%
\end{equation}
for polynomials $q_{j}\left(  x\right)  =a_{0j}+a_{2j}x^{2}+a_{4j}x^{4}$,
$j=1,2,3$. The constants $a_{ij}$ forming $\mathbf{a}$ are determined by the
constraints in terms of the $\mu_{k}$, $\kappa_{k}$ and $\beta^{2}$, which are
then optimally chosen to minimize the loss. It is simpler however to choose
$\mathbf{a}$ directly, to minimize $\mathcal{L}_{\nu}\left(  \xi
|\sigma\right)  $ over all densities of the form (\ref{form}), subject to
$\int_{-1}^{1}m(x;\mathbf{a})dx=1$.

\section*{Acknowledgements}

This work has been supported by the Natural Sciences and Engineering Research
Council of Canada.

\section*{References}

\begin{description}
\item Daemi, M., and Wiens, D. P. (2013), \textquotedblleft Techniques for the
Construction of Robust Regression Designs,\textquotedblright\ \textit{The
Canadian Journal of Statistics}, 41, 679 - 695.

\item Knight, K. (1998), \textquotedblleft Limiting Distributions for $l_{1}$
Estimators Under General Conditions,\textquotedblright\ \textit{Annals of
Statistics}, 26, 755-770.

\item Pollard, D. (1991), \textquotedblleft Asymptotics for Least Absolute
Deviation Regression Estimators,\textquotedblright\ \textit{Econometric
Theory}, 7, 186-199.
\end{description}

\end{document}